\documentclass[useAMS,usenatbib,twocolumn]{mnras}
\usepackage{natbib}
\usepackage{graphicx}
\usepackage{rotate}
\usepackage{hyperref}
\usepackage{breakurl}
\newcommand{\redm}{redMaPPer }

\newcommand{\eg}{{\it e.g.,}}

\newcommand{\cat}{{\sc s82-mgc}}   % product name

\newcommand{\redmapper}{{\it redMaPPer}}  

\begin{document}
  
%-------- TITLE  ---------------------

 \title[Stellar Mass Completeness of BOSS]{The Stripe 82 Massive
   Galaxy Project II: Stellar Mass Completeness
   of Spectroscopic Galaxy Samples from the Baryon Oscillation
   Spectroscopic Survey}

 %% LaTeX will automatically break titles if they run longer than
 %% one line. However, you may use \\ to force a line break if
 %% you desire.

%-------- AUTHORS  ---------------------

 %% Use \author, \affil, and the \and command to format
 %% author and affiliation information.
 %% Note that \email has replaced the old \authoremail command
 %% from AASTeX v4.0. You can use \email to mark an email address
 %% anywhere in the paper, not just in the front matter.
 %% As in the title, use \\ to force line breaks.

 \author[Leauthaud et al.]  {Alexie Leauthaud$^1$, Kevin Bundy$^1$,
   Shun Saito$^1$, Jeremy Tinker$^2$, Claudia Maraston$^{3}$,
   \newauthor
   Rita Tojeiro$^{4}$, Song Huang$^1$, Joel R. Brownstein$^5$, Donald
   P. Schneider$^{6,7}$,\newauthor
 Daniel Thomas$^3$\\
   $^1$Kavli IPMU (WPI), UTIAS, The University of Tokyo, Kashiwa,
   Chiba
   277-8583, Japan\\
   $^2$Center for Cosmology and Particle Physics, Department of
   Physics,
   New York University\\
   $^3$ Institute of Cosmology and Gravitation, University of Portsmouth,
Portsmouth, PO1 3FX, UK\\
   $^4$School of Physics and Astronomy, University of St Andrews, St
   Andrews, KY16 9SS, UK\\
    $^5$Department of Physics and Astronomy, University of Utah, 115 S
  1400 E, Salt Lake City, UT 84112, USA\\
$^6$ Department of Astronomy and Astrophysics, The Pennsylvania State University,
  University Park, PA 16802\\
  $^7$ Institute for Gravitation and the Cosmos, The Pennsylvania State University,
  University Park, PA 16802}
\maketitle
\label{firstpage}

%-------- ABSTRACT  ---------------------
  
%\gtrapprox

\begin{abstract} The Baryon Oscillation Spectroscopic Survey (BOSS)
  has collected spectra for over one million galaxies at $0.15<z<0.7$
  over a volume of 15.3 Gpc$^3$ (9,376 deg$^2$) -- providing us an
  opportunity to study the most massive galaxy populations with
  vanishing sample variance. However, BOSS samples are selected via
  complex color cuts that are optimized for cosmology studies, not
  galaxy science. In this paper, we supplement BOSS samples with
  photometric redshifts from the Stripe 82 Massive Galaxy Catalog and
  measure the total galaxy stellar mass function (SMF) at $z\sim0.3$
  and $z\sim0.55$. With the total SMF in hand, we characterize the
  stellar mass completeness of BOSS samples. The high-redshift CMASS
  (``constant mass'') sample is significantly impacted by mass
  incompleteness and is 80\% complete at $\log_{10}(M_*/M_{\odot}) >
  11.6$ only in the narrow redshift range $z=[0.51,0.61]$. The low
  redshift LOWZ sample is 80\% complete at $\log_{10}(M_*/M_{\odot}) >
  11.6$ for $z=[0.15,0.43]$. To construct mass complete samples at
  lower masses, spectroscopic samples need to be significantly
  supplemented by photometric redshifts. This work will enable future
  studies to better utilize the BOSS samples for galaxy-formation
  science.\end{abstract}

 
%-------- KEY WORDS  ---------------------

 %% Keywords should appear after the \end{abstract} command. The uncommented
 %% example has been keyed in ApJ style. See the instructions to authors
 %% for the journal to which you are submitting your paper to determine
 %% what keyword punctuation is appropriate.
 
%\keywords{cosmology: observations -- gravitational lensing -- large-scale
%structure of Universe}
 
 %% Authors who wish to have the most important objects in their paper
 %% linked in the electronic edition to a data center may do so by tagging
 %% their objects with \objectname{} or \object{}.  Each macro takes the
 %% object name as its required argument. The optional, square-bracket 
 %% argument should be used in cases where the data center identification
 %% differs from what is to be printed in the paper.  The text appearing 
 %% in curly braces is what will appear in print in the published paper. 
 %% If the object name is recognized by the data centers, it will be linked
 %% in the electronic edition to the object data available at the data centers

%%%%%%%%%%%%%%%%%%%%%%%%%%%%%%%%%%%%%%%%%%%%%%%%%%%%%%%%%%%%%%%%%%%%%%%%%%%%%%
%     INTRODUCTION
%%%%%%%%%%%%%%%%%%%%%%%%%%%%%%%%%%%%%%%%%%%%%%%%%%%%%%%%%%%%%%%%%%%%%%%%%%%%%%

\section{Introduction}

There is tremendous interest in constraining the size, stellar mass,
and halo mass evolution of the most massive galaxies in the
universe \citep[][to cite a few recent
examples]{Tojeiro:2012,Tal:2013,Maraston:2013,
  Beifiori:2014,Marchesini:2014,van-de-Sande:2015,Marsan:2015}. The
evolution of these properties places strong constraints on models of
galaxy formation which traditionally have difficulty reproducing
observed trends such as the amplitude of the stellar mass function at
the highest masses \citep[\eg][]{Benson:2003,Maraston:2013,
  Knebe:2015}, although much progress has been made in the past few
years \citep[\eg][]{Furlong:2014,Benson:2014a}.

From an observational standpoint, spectroscopic samples of massive
galaxies present key advantages over photometric samples. For example,
errors in stellar mass estimates are reduced for spectroscopic samples
compared to photometric redshift samples. Typical 5-band photometric
redshifts at $z\sim0.5$ have an error of $\sigma_{z}=0.03$ to
$\sigma_{z}=0.05$ even for the most massive galaxies. This error
translates into a stellar mass uncertainty of $\sim$0.1 dex which may
dominate the total stellar mass error budget. Spectroscopic samples of
massive galaxies are also key in order to perform accurate
measurements of clustering and/or galaxy-galaxy lensing which place
tight constraints on the galaxy-halo connection
\citep[\eg][]{Mandelbaum:2006c,Leauthaud:2012,
  Coupon:2015,Zu:2015}. Finally, the spectra themselves contain key
information and can be used to constrain stellar ages, star formation
histories (SFHs), dust extinctions and stellar velocity dispersions
\citep[\eg][]{Chen:2012,Thomas:2013}.

For these reasons, from a galaxy formation perspective, large samples
of massive galaxies with spectroscopic redshifts over a wide redshift
range are highly desirable. Surveys such as zCOSMOS
\citep[][]{Lilly:2007}, VVDS \citep[][]{Le-Fevre:2004,Le-Fevre:2015},
DEEP2 \citep[][]{Newman:2013a}, PRIMUS \citep[][]{Coil:2011}, and
VIPERS \citep[][]{Guzzo:2014} provide spectroscopic samples that probe
the $0.2<z<1.0$ universe and complement the Sloan Digital Sky Survey
\citep[SDSS,][]{York:2000} main sample at $z=0.1$
\citep[][]{Strauss:2002}. These higher redshift surveys, however,
still cover relatively small areas ranging from a few square degrees
to a few tens of square degrees (\eg VIPERS covers 24 deg$^2$). The
volumes probed by these surveys are insufficient to provide
statistically significant samples of the most massive galaxies
($\log(M_*/M_{\odot})>11.5$) which have low number densities
($\overline{n}\sim 2\times10^{-5}$ Mpc$^{-3}$). For example, the
highest mass bin at $0.48<z<0.74$ in the \citet[][]{Leauthaud:2012}
COSMOS analysis of the stellar-to-halo mass relation only contains 300
galaxies at $\log(M_*/M_{\odot})>11.29$, only 71 of which have masses
greater than $\log(M_*/M_{\odot})>11.5$. The highest mass bin in the
\citet[][]{Coupon:2015} analysis, which covers a wider area (23.1
deg$^2$), contains 6326 galaxies at $0.5<z<1.0$ and
$\log(M_*/M_{\odot})>11.2$. However, this bin only contains 498
galaxies at $\log(M_*/M_{\odot})>11.5$, of which only 234 have a
secure spectroscopic redshift from the VIPERS, VVDS, or PRIMUS
surveys. Because of these small samples sizes, studies of the
stellar-to-halo mass relation, for example, remain poorly constrained
at the very high-mass end.

An exciting opportunity is the Baryon Oscillation Spectroscopic Survey
\citep[BOSS,][]{Eisenstein:2011, Dawson:2013}, which, with the final
DR12 data release \citep[][]{Alam:2015}, has collected spectra for
more than one million galaxies massive galaxies
($\log(M_*/M_{\odot})>11.0$) at $0.15<z<0.7$ over a volume of 15.3
Gpc$^3$ (9,376 deg$^2$), providing the potential to study the most
massive galaxy populations with vanishing sample variance.  However,
the sample selections of BAO surveys involve complex color cuts that
are optimized for cosmology studies, not galaxy science. While a
number of BOSS galaxy studies have been published \citep[][to cite a
few]{Tojeiro:2012,Maraston:2013,Guo:2013,Beifiori:2014,
  Guo:2014,Montero-Dorta:2014,Montero-Dorta:2015, Reid:2014,
  Guo:2015}, the stellar mass completeness of the BOSS samples remains
poorly understood.

In this paper, the second in a series, we use a new compilation of
wide-field survey data, the Stripe 82 Massive Galaxy Catalog (\cat{}),
to address this problem.  Paper I (Bundy et al., in preparation)
describes the construction of \cat{}, which matches the 2 magnitudes
deeper ``SDSS Coadd'' optical photometry \citep[][]{Annis:2011} in the
equatorial Stripe 82 with the Large Area Survey (LAS) near-IR
photometry from the UKIRT Infrared Deep Sky Survey
\citep[UKIDSS,][]{Lawrence:2007}.  Supplementing with a variety of
photometric redshifts, \cat{} enables near-IR based stellar mass
estimates for complete samples with $\log(M_*/M_{\odot})>11.2$ and
$z < 0.7$.

In this paper, we use the \cat{} to investigate the stellar mass
completeness of the two main BOSS spectroscopic samples, the LOWZ
sample at $0.15<z<0.43$ and the CMASS (``Constant Mass'') sample at
$0.43<z<0.7$. Our characterization of the completeness is made with
respect to a measurement of the galaxy stellar mass function using the
\cat{}.  The evolution of the mass function, a detailed study of
potential biases, and implications for galaxy growth are discussed in
Paper III (Bundy et al., in preparation).  Here, we provide convenient
fitting formulae which can be used to estimate the completeness of
each of the BOSS samples as a function of stellar mass and redshift
(Sections \ref{sect_cmass_comp} and \ref{sect_lowz_comp}).

This work is complementary to the analysis performed by
\citet[][]{Montero-Dorta:2014} who studied the magnitude and color
completeness of the high-redshift BOSS CMASS sample. This paper
focuses on stellar mass completeness for the full CMASS and LOWZ
samples, an aspect that is not addressed in
\citet[][]{Montero-Dorta:2014}.

Upcoming surveys such as the Hyper Suprime Cam
survey\footnote{\url{http://www.naoj.org/Projects/HSC/HSCProject.html}}
(HSC) and the Euclid survey \citep[][]{Laureijs:2011} will be able to
use photometric redshifts to supplement spectroscopic samples and to
construct mass limited samples over wider redshift and mass ranges
than using spectroscopic samples alone. We characterize the level to
which spectroscopic samples need to be supplemented by photometric
redshifts as a function of mass and redshift (see Section
\ref{mstarsample}).

Finally, a better understanding of the BOSS selection functions will
also enable the construction of improved mock catalogs that are
critical to investigating the link between galaxies and their host
halos. Our companion paper (Saito et al., in preparation) presents
improved mock catalogs that account for the stellar mass completeness
of the BOSS CMASS sample as a function of redshift.

The layout of this paper is as follows. Section \ref{data} presents
the data used in this paper. Section \ref{mstarsample} discusses the
characteristics of of sample. Section \ref{colorcuts} broadly
describes the effects of the BOSS sample selection. Section
\ref{totalsmf} presents our estimate of the total stellar mass
function as a function of redshift. Section \ref{comp} describes our
completeness estimates for CMASS and LOWZ. Section \ref{previous_work}
presents mass completeness estimates for several previous studies that
used BOSS data. Finally, our summary and conclusions are presented in
Section \ref{conclusions}. We assume a $\Lambda$CDM cosmology with
$\Omega_{\rm m}=0.274$, $H_0=70$ km~s$^{-1}$~Mpc$^{-1}$. Stellar mass
is noted $M_{*}$ and has been derived using a Chabrier Initial Mass
Function \citep[IMF,][]{Chabrier:2003}.

\section{Data}\label{data}

\subsection{BOSS Spectroscopic Samples}\label{bossdata}

% Anderson 2012 DR9
% Anderson 2014 DR10/11

BOSS is a spectroscopic survey of 1.5 million galaxies over 10,000
deg$^2$ that was conducted as part of the SDSS-III program
\citep[][]{Eisenstein:2011} on the 2.5 m aperture Sloan Foundation
Telescope at Apache Point Observatory \citep[][]{Gunn:1998,
  Gunn:2006}. A general overview of the BOSS survey can be found in
\citet[]{Dawson:2013}, the BOSS spectrographs are described in
\citet[]{Smee:2013}, and the BOSS pipeline is described in
\citet[]{Bolton:2012}. BOSS galaxies were selected from Data Release 8
\citep[DR8,][]{Aihara:2011} {\it ugriz} imaging
\citep[][]{Fukugita:1996} using a series of color-magnitude cuts
motivated by the \citet{Maraston:2009} LRG model. This is a passive
template mainly dominated by a metal-rich population but also includes
a small metal-poor population ($3\%$~by mass) to mimic observed
metallicity gradients of local massive elliptical galaxies. This
template was shown to fit the observed-frame Sloan colors of 2-SLAQ
galaxies at redshift 0.4-0.6 better than models which include star
formation \citep[][]{Wake:2006}. This conclusion was also
independently confirmed by \citet[][]{Montero-Dorta:2015}. The choice
of a passive template was motivated by the intention of selecting the
most massive and passive galaxies for BAO studies.

The BOSS selection uses the following set of color criteria:

\begin{eqnarray}
c_\parallel &=& 0.7(g_{\rm mod}-r_{\rm mod}) + 1.2[(r_{\rm mod}-i_{\rm
  mod}) - 0.18] \\
 c_\perp &=& (r_{\rm mod}-i_{\rm mod}) - (g_{\rm mod}-r_{\rm mod})/4 - 0.18\\
d_\perp &=& (r_{\rm mod}-i_{\rm mod})-(g_{\rm mod}-r_{\rm mod})/8.0
\end{eqnarray} 

The subscript ``mod'' denotes model magnitudes, which are derived by
adopting the better fitting luminosity profile between a de
Vaucouleurs and an exponential luminosity profile in the r-band
\citep[][]{Stoughton:2002}. The subscript ``cmod'' denotes composite
model magnitudes, which are calculated from the best-fitting linear
combination of a de Vaucouleurs and an exponential luminosity profile
\citep[][]{Abazajian:2004}. PSF magnitudes are denoted with the
subscript ``psf''. BOSS color cuts are computed using model
magnitudes, whereas magnitude cuts are computed using cmodel
magnitudes. All magnitudes are corrected for Galactic extinction using
the dust maps of \citet{Schlegel:1998}.

% ----- LOWZ  --------

BOSS targeted two primary galaxy samples: the LOWZ sample at
$0.15<z<0.43$ and the CMASS sample at $0.43<z<0.7$. The LOWZ sample is
an extension of the SDSS I/II Luminous Red Galaxy (LRG) sample
\citep[][]{Eisenstein:2001} to fainter magnitudes and is defined
according to the following selection criteria:

\begin{eqnarray}
|c_\perp| &<& 0.2\label{lowz_cperp}\\
r_{\rm cmod} &<& 13.6 + c_\parallel/0.3\label{lowz_sliding}\\
16 < r_{\rm cmod} &<& 19.6\label{lowz_flux_limit}\\
r_{\rm psf} - r_{\rm cmod} &>& 0.3\label{star_gal}
\end{eqnarray}

Equation \ref{lowz_cperp} sets the color boundaries of the sample;
equation \ref{lowz_sliding} is a sliding magnitude cut which selects
the brightest galaxies at each redshift; equation
\ref{lowz_flux_limit} corresponds to the bright and faint limits and
equation \ref{star_gal} is to separate galaxies from stars. In a
similar fashion to the SDSS I/II Luminous Red Galaxy (LRG) sample, the
LOWZ selection primarily selects red galaxies. Over most of the BOSS
footprint, roughly one third of the LOWZ sample has a spectrum from
SDSS-II (these objects were not re-observed by BOSS). Galaxies with a
spectrum that pre-dates the BOSS survey will be referred to
interchangeably either as ``Legacy objects'' or ``SDSS KNOWN''.

% ----- CMASS  --------

The CMASS sample targets galaxies at higher redshifts with a surface
density of roughly 120 deg$^{-2}$. CMASS targets are selected from
SDSS DR8 imaging according to the following cuts:

\begin{eqnarray}
|d_\perp| &>& 0.55\label{cmass_dperp}\\
i_{\rm cmod} &<& 19.86 +1.6(d_\perp-0.8)\label{cmass_sliding}\\
17.5 &<& i_{\rm cmod} < 19.9\label{cmass_flux_limit}\\
r_{\rm mod} - i_{\rm mod} &<& 2\\
i_{\rm fib2}&<&21.5
\end{eqnarray}

\noindent where $i_{\rm fib2}$ is the estimated i-band magnitude in a
2\arcsec\ aperture diameter assuming 2\arcsec\ seeing. Star-galaxy
separation on CMASS targets is performed via:

\begin{eqnarray}
i_{\rm psf} - i_{\rm mod} &>&0.2+0.2(20.0-i_{\rm mod})\\
z_{\rm psf} - z_{\rm mod} &>&9.125-0.46z_{\rm mod}.
\end{eqnarray}

In addition to CMASS, BOSS also targeted a smaller ancillary sample
known as ``CMASS SPARSE'' that was designed to test the impact of the
color-magnitude cuts. The SPARSE sample includes fainter and bluer
galaxies by extending the sliding cut to:

\begin{equation}\label{cmass_sliding_sparse}
i_{\rm cmod} < 20.14+1.6(d_\perp-0.8). 
\end{equation}

The SPARSE region (contained between the fiducial CMASS sliding cut
and the SPARSE sliding cut) is randomly down-sampled to yield about 5
objects per square degree. Beyond highlighting the SPARSE region in
certain figures, we do not utilize this sample in this paper. Instead,
our analysis relies on photometric redshifts to probe the full massive
galaxy population that extends beyond the BOSS color boundaries (see
Section \ref{s82}).

The CMASS sample was originally designed to loosely follow a constant
stellar mass cut at $0.4<z<0.6$ (see Figure~1 in
\citealt{Maraston:2013}) and to allow for a wider range of galaxy
colors than either the SDSS-II LRG or the LOWZ samples. Using
high-resolution HST/ACS imaging, \citet{Masters:2011} show that
roughly 26\% of the CMASS galaxies in the COSMOS survey are
morphologically classified as late-types (with an observed color of
$g-i<2.35$). Using a maximum likelihood approach that accounts for
photometric errors as well as the CMASS selection cuts,
\citet{Montero-Dorta:2014} estimate that 37\% of CMASS object may
intrinsically belong to the blue cloud.

The CMASS sample is thought to be more complete at higher stellar
masses than the SDSS-II LRG and LOWZ samples which are color-selected
\citep[][]{Anderson:2014}. However, we will demonstrate in Section
\ref{sect_lowz_comp} that the opposite is in fact true and that in
certain redshift ranges, the LOWZ sample is more complete is terms of
stellar mass than CMASS (see Section \ref{sect_lowz_comp}).

% ----- SAMPLES --------

In this paper, we use the internal DR10 BOSS data release
\citep[][]{Ahn:2014}. However, our analysis is limited to the Stripe
82 region of the BOSS survey (described in Section \ref{s82}) which
was observed in the early phases of the BOSS survey. Because DR11 and
DR12 provide no new data in Stripe 82, our analysis and conclusion
would be identical using these later data releases. The main samples
and catalogs that are used in this paper are:

\begin{enumerate}
\item The original catalog that was used to target BOSS galaxies. From
  this catalog, we extract the original set of LOWZ and CMASS
  targets. We will refer to the original target samples as
  ``TAR\_LOWZ'' and ``TAR\_CMASS''. Galaxies that were part of the
  target catalog, but did not obtain a fiber due to fiber-collisions
  are designated as ``FIBER\_COLLIDED\_LOWZ'' and
  ``FIBER\_COLLIDED\_CMASS''.
%NOTE: LSS catalog does not include SPARSE
\item The BOSS DR10 large-scale structure (LSS) catalog created by the
  BOSS galaxy clustering working group that was used in the
  \citet[][]{Anderson:2014} analysis. We only give a brief description
  of this catalog and refer the reader to \citet[][]{Anderson:2014}
  for additional details. This catalog is constructed by trimming the
  BOSS redshift file (the ``SpALL'' file) by the BOSS mask. Legacy
  objects are added to the catalog (Legacy objects are not contained
  in the SpALL file). Finally, Legacy objects are down-sampled in each
  sector to match the BOSS completeness. The LSS catalog contains the
  flag ``IMATCH'' which indicates galaxies with BOSS redshifts
  (IMATCH$=1$) and galaxies with Legacy redshifts
  (IMATCH$=2$). Galaxies from the LSS catalogs will be referred to as
  ``LSS\_LOWZ'' and ``LSS\_CMASS''\footnote{Note that we do not apply
    the large-scale systematic weights that are applied to the BOSS
    clustering analyses.}.
\item The BOSS stellar mass catalog from the Portsmouth group
  \citep[][]{Maraston:2013}.
\end{enumerate}

Further details and the url's of these catalogs are provided in
the Appendix.

\subsection{The Stripe 82 Massive Galaxy Catalog}\label{s82}

A key data set for this paper comes from observations of the Stripe 82
region along the celestial equator in the region of the southern
galactic sky---a narrow, but deeper subset of the SDSS survey
area---for which it is possible to construct a galaxy sample with a
well understood completeness function. Stripe 82 is critical for this
paper for two reasons. First, it was the subject of repeat imaging
campaigns in SDSS, especially by the SDSS Supernova Survey
\citep[][]{Frieman:2008a}. These data have been combined into the
``SDSS Coadd'' by \citet{Annis:2011} and reach roughly 2 magnitudes
deeper than the single epoch SDSS imaging with a 5$\sigma$ detection
limit of $r \sim 22.5$. This added depth is critical for obtaining
reliable photometric redshifts (photo-z’s) for massive galaxies that
can be used to supplement the color-selected BOSS samples out to
$z\sim0.7$. Second, this region was also imaged by the UKIRT Infrared
Deep Sky Survey \citep[UKIDSS,][]{Lawrence:2007} which provides
near-IR photometry for ensuring robust stellar mass estimates.

The Stripe 82 Massive Galaxy Catalog (\cat{}) combines these data sets
and delivers matched $ugrizYJHK$ photometry using catalog-level
synthetic aperture photometry \citep{Bundy:2012aa}.  Galaxies are
separated from stars following \citet{Baldry:2010} and via a
combination of light profile and color cuts.  Redshift information is
provided by a number of sources including all SDSS spectroscopic
redshifts available in DR10 and several photometric redshift
estimates.  Among these are neural-network-based photo-z’s from
\citet{Reis:2012} and iterative template-fitting estimates from the
red-sequence Matched-filter Probabilistic Percolation
\citep[\redmapper{},][]{Rykoff:2014}.  The typical photo-z precision
is $\sigma_z\simeq 0.03$ at $z\leq 0.7$.

A redshift estimator, $z_{\rm best}$, is built for {\sc s82-mgc} that
employs spectroscopic redshifts whenever possible and chooses among
three photometric-based redshift estimates otherwise.  In order of
priority, $z_{\rm best}$ is set to:

\begin{enumerate}
\item $z_{\rm spec}$: spectroscopy from SDSS-II or from BOSS.
\item $z_{\lambda}$: for cluster members.
\item $z_{\rm red}$: for field early type galaxies.
\item $z_{\rm phot}$: photometric redshift from \citet{Reis:2012}.
\end{enumerate}

The scatter and bias in the photometric-based redshift estimators are
characterized in Paper I.

Mangle \citep[][]{Swanson:2008} and custom software is used to apply
geometric masking of regions with imaging artifacts in the SDSS coadds
and UKIDSS imaging and areas around bright stars. The BOSS acceptance
mask as well as rejection masks for collisions with the plate post and
potentially higher priority BOSS targets are also applied\footnote{For details on BOSS masks, see: \url{
    http://www.sdss3.org/dr10/tutorials/lss\_galaxy}.}.

Fiducial stellar mass estimates for the {\sc s82-mgc} are described in
Paper I and compared to other publicly available estimates from the
BOSS survey.  A more detailed examination of $M_*$ estimators and
potential biases is presented in Paper III.  Briefly, the fiducial
$M_*$ estimates are the result of the SED-fitting code described in
\citet[][]{Bundy:2010} applied to the SDSS+UKIDSS PSF-matched
photometry and the defined $z_{\rm best}$.  Stellar population
templates are derived using BC03 models \citep{Bruzual:2003} and the
STELIB empirical library \citep{Le-Borgne:2003} and assuming a
\citet{Chabrier:2003} Initial Mass Function (IMF). For a prior grid of
SED templates spanning a range of ages (restricted to be less than the
age of the universe at the redshift of the galaxy), metallicities,
dust extinction, and exponential $\tau$ star formation histories, a mass
probability distribution is obtained by scaling the model $M/L$ ratios
by the inferred luminosity in the observed K-band. The median of this
distribution is taken as the stellar mass estimate.

Starting from the \cat{} parent catalog, we select the ``UKWIDE''
sub-sample (see Paper I) by rejecting sources in masked regions and
requiring all classified galaxies be observed on UKIDSS-LAS frames
with 5$\sigma$ detection limits deeper than {\it
  YJHK}$=$[20.3,20.0,19.6,19.5] on the AB system of
\citet[][]{Oke:1983}. The final UKWIDE sample spans an area of 139.4
deg$^2$.

%%%%%%%%%%%%%%%%%%%%%%%%%%%%%%%%%%%%%%%%%%%%%%%%%%%%%%%%%%%%%%%%%%%%%%%%%%%%%%
%     MASSIVE GAL SAMPLE
%%%%%%%%%%%%%%%%%%%%%%%%%%%%%%%%%%%%%%%%%%%%%%%%%%%%%%%%%%%%%%%%%%%%%%%%%%%%%%

\section{Sample Characteristics}\label{mstarsample}

\begin{figure}
\begin{center}
\includegraphics[width=8.6cm]{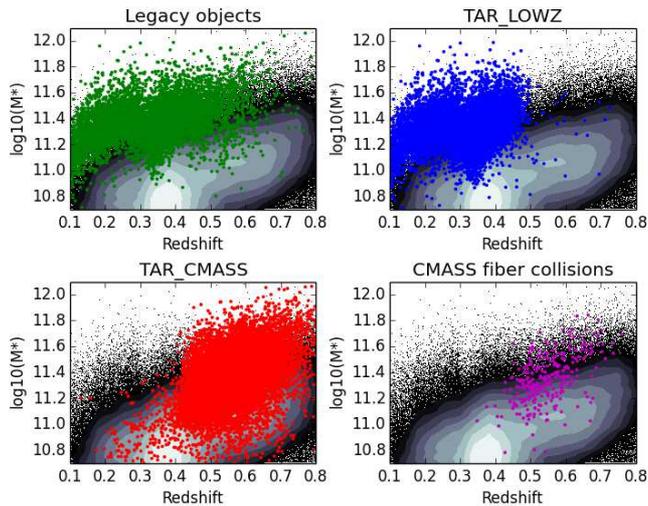}
\caption{Mass and redshift distributions of several galaxy samples
  considered in this paper (indicated by colored symbols). The
  underlying black points and grey contours show all galaxies from the
  {\sc s82-mgc} as a function of mass and redshift. The {\sc s82-mgc}
  catalog extends to lower mass limits compared to the BOSS
  spectroscopic samples.}
\label{mass_z}
\end{center}
\end{figure}

Figure \ref{mass_z} displays the stellar mass and redshift
distributions for the main galaxy samples considered in this paper
compared to the full galaxy population from the {\sc s82-mgc}. The
{\sc s82-mgc} catalog extends to lower mass limits compared to the
BOSS samples and is mass complete to $\log_{10}(M_∗/M_{\odot})=11.2$
at $z=0.7$ (Paper I).

% /Users/alexie/Work/Boss/Completeness/Pycode/mass_z_plot.py

For the LOWZ sample, because of a targeting error\footnote{See
  \citealt{Parejko:2013} and \url{http://www.sdss3.org/
    dr9/algorithms/boss\_galaxy\_ts.php}} related to star-galaxy
separation in the early phase of the BOSS survey, a cut of
TILE$>$10324 must be applied in order to select a uniform sample of
LOWZ galaxies. However, the Stripe 82 region was unaffected by this
error (see Figure 2 in \citealt{Parejko:2013}) and this cut is
unnecessary for the {\sc s82-mgc}.

\begin{figure}
\begin{center}
\includegraphics[width=8.5cm]{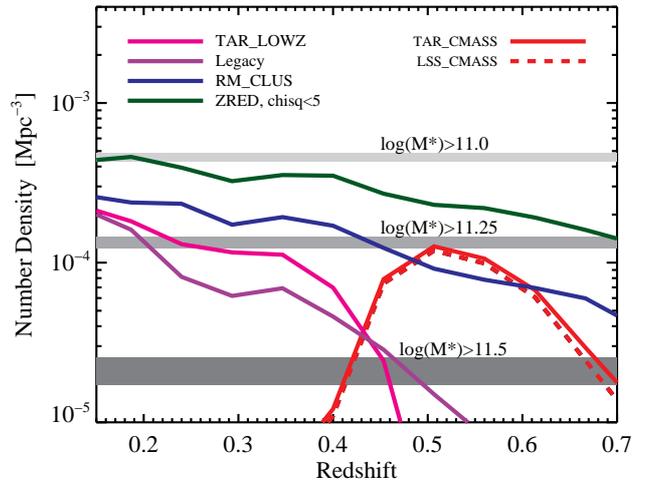}
\caption{Number density as a function of redshift for several galaxy
  samples considered in this paper. The magenta line corresponds to
  LOWZ targets (``TAR\_LOWZ''). The purple line corresponds to Legacy
  SDSS-II LRGs (``Legacy''). The blue line corresponds to \redm
  cluster members \citep[``RM\_CLUS'',][]{Rykoff:2014}. The green line
  corresponds to galaxies with $z_{\rm red}$ and $\chi^2_{\rm red}<5$
  (``ZRED''). The solid red line corresponds to CMASS targets
  (``TAR\_CMASS'') and the dashed red line corresponds to CMASS galaxies from the LSS
  catalog (``LSS\_CMASS''). Grey dashed lines indicate the number densities of stellar mass
  threshold samples as a function of redshift estimated from the {\sc
    s82-mgc}.}
\label{nden_fig}
\end{center}
\end{figure}

Figure \ref{nden_fig} compares the number densities of various samples
(e.g., Legacy, LOWZ, CMASS) to the estimated number density of stellar
mass threshold samples from the {\sc s82-mgc}.

\begin{figure*}
\begin{center}
\includegraphics[width=16cm]{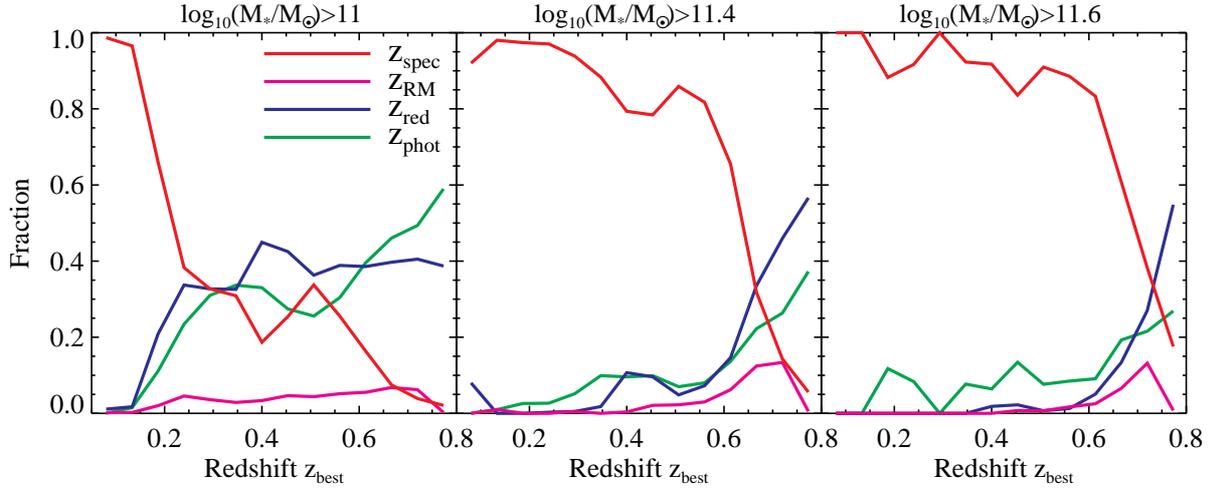}
\caption{Origin of redshifts that contribute to $z_{\rm best}$ as a
  function of stellar mass. The solid red line indicates the fraction
  of galaxies with spectroscopic redshifts and with
  $\log_{10}(M_*/M_{\odot})>11$ (left panel),
  $\log_{10}(M_*/M_{\odot})>11.4$ (middle panel), and
  $\log_{10}(M_*/M_{\odot})>11.6$ (right panel) as a function of
  redshift. At $z< 0.61$, it is possible to construct mass-limited
  samples with $\log_{10}(M_∗/M_{\odot}) > 11.6$ for which more than
  80\% of the sample has a spectroscopic redshift. At
  $\log_{10}(M_∗/M_{\odot}) > 11.0$, however, spectroscopic samples
  must be significantly supplemented by photometric redshifts, except
  at low redshifts ($z< 0.1$).}
\label{zbest_frac}
\end{center}
\end{figure*}

Large imaging surveys that will overlap with the BOSS footprint such
as the HSC and Euclid surveys will be able to build mass limited
samples using a combination of spectroscopic redshifts and
photometric-based estimators using a strategy similar to the one we
have used here. The ratio $N_{\rm spec}/N_{\rm phot}$ represents a
useful quantity when considering trade-offs between the mass limits of
such samples and errors introduced by supplementing spectroscopic
samples with photometric redshifts. Figure \ref{zbest_frac} presents
the origin of redshifts that contribute to $z_{\rm best}$ for three
stellar mass thresholds. At $z<0.61$, it is possible to construct
mass-limited samples with $\log_{10}(M_∗/M_{\odot}) > 11.6$ for which
more than 80\% of the sample has a spectroscopic redshift. At
$\log_{10}(M_∗/M_{\odot}) > 11.0$, however, spectroscopic samples must
be significantly supplemented by photometric redshifts, except at low
redshifts ($z< 0.1$).

\begin{figure*}
\begin{center}
\includegraphics[width=16cm]{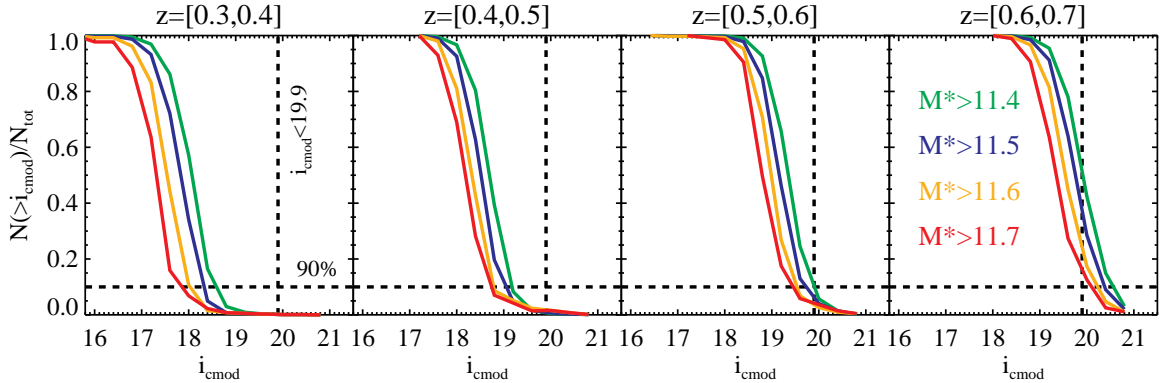}
\caption{Cumulative magnitude distribution of stellar-mass threshold
  samples as a function of redshift. The dashed horizontal lines shows
  the 90\% completeness limit. The dashed vertical line corresponds to
  $i_{\rm cmod}=20$. One of the cuts that defines the CMASS samples is
  a magnitude cut at $i_{\rm cmod}< 19.9$. This figure
  demonstrates, for example, that galaxy samples with
  $i_{\rm cmod}< 20$ are $90\%$ complete for $\log_{10}(M_*)=11.7$
  at $z=0.65$.}
\label{mag_limits}
\end{center}
\end{figure*}

Finally, another question of interest is to determine how faint a
spectroscopic survey needs to reach in order to probe a certain mass
limit at a given redshift. Figure \ref{mag_limits} presents the
cumulative $i_{\rm cmod}$ magnitude distribution of stellar mass
threshold samples as a function of redshift. This figure demonstrates
that galaxy samples with $i_{\rm cmod}< 20$ are roughly $90\%$
complete for $\log_{10}(M_*)=11.7$ at $z=0.65$. For comparison, CMASS
includes a cut at $i_{\rm cmod} < 19.9$ which impacts the completeness at
higher redshifts. A more in-depth study of the impact of the CMASS
cuts is provided in the following section.

\section{Effects of Target Selection on BOSS Galaxy Samples}\label{colorcuts}

% \begin{equation}
% b_{1000}=\frac{\int_{t}^{t+1 Gyr}SFR(t)dt}{\int_{t}^{T}SFR(t)dt}
% \end{equation}

We begin with a broad investigation of the impact of the BOSS target
selection on the BOSS galaxy samples as a function of redshift,
stellar mass, and color. To represent galaxy colors we use an estimate
of the birth parameter, $b_{1000}$, which is the ratio of the average
star formation rate within the previous 1 Gyr to the star formation
rate averaged over the galaxy's history.  In this work, we use
$b_{1000}$ estimates from the KCORRECT package
\citep[][]{Blanton:2007} which are computed from an SED fit of a
linear combination of stellar population templates from
\citet{Bruzual:2003} (hereafter, BC03). As for $M_*$ estimates,
$b_{1000}$ depends on the assumed models and priors used to the fit
the observed SEDs.  It is a rough estimate of recent star formation
that is based on more than a single optical color (such as $g-r$) and
which takes advantage of the added constraints on dust afforded by the
near-IR photometry.

In the {\sc s82-mgc}, galaxies at $z\sim0.55$ and $M_*\sim11.5$ have
$b_{1000}$ values as high as 0.7, suggesting an occasional high rate
of recent star formation, but the vast majority are peaked near
$b_{1000} = 0$ as expected for an old, passively evolving stellar
population. Paper I explores in greater detail how $b_{1000}$ better
separates red-sequence galaxies with ongoing dusty star formation from
those with truly passive populations.

\begin{figure*}
\begin{center}
\includegraphics[width=14cm]{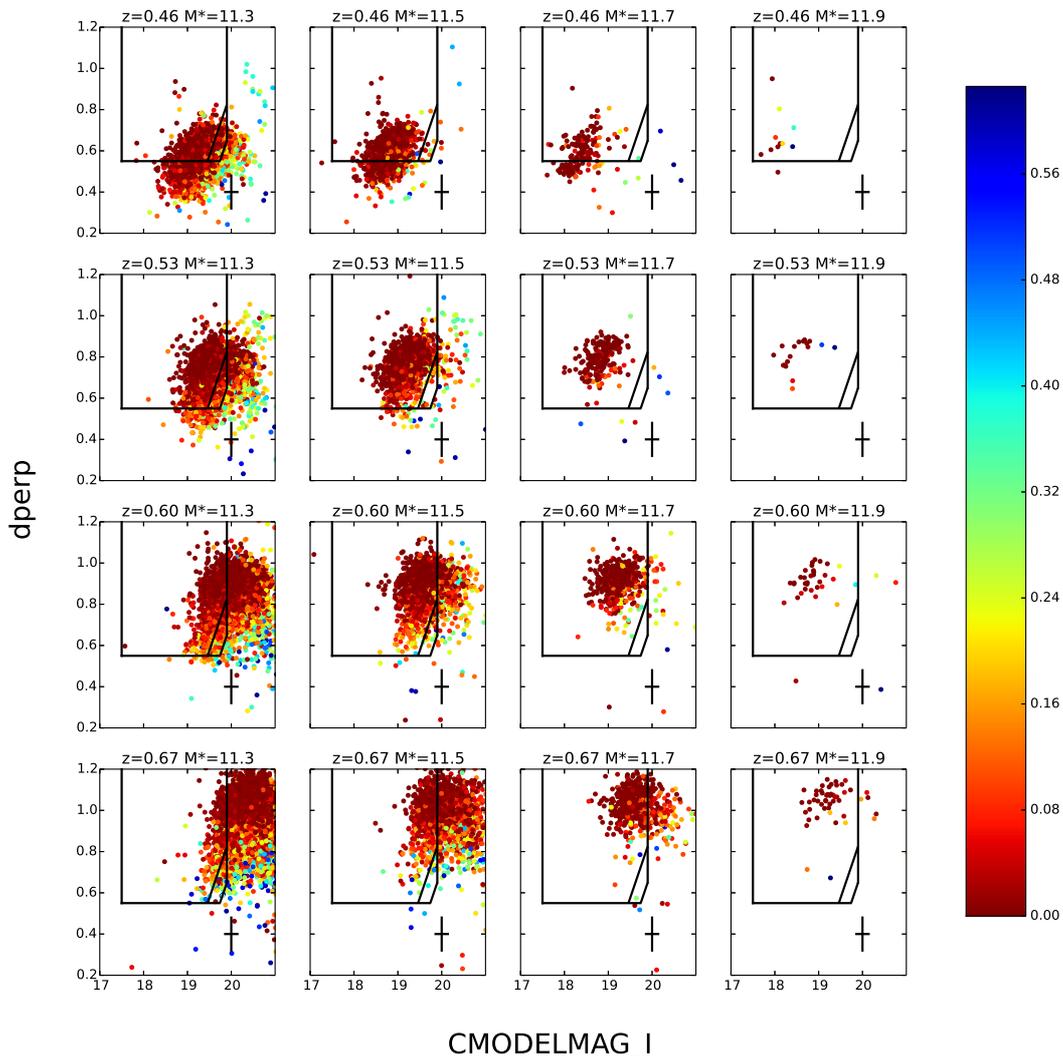}
\caption{Galaxies from the {\sc s82-mgc} in the CMASS redshift range
  as a function of $i_{\rm cmod}$ and $d_{\rm perp}$. Galaxies are
  color coded according to their birth parameter $b_{1000}$: larger
  values of $b_{1000}$ (bluer colors) indicate galaxies that have
  experienced larger amounts of star formation in the past
  Gyr. Vertical black lines indicate the CMASS flux limits (equation
  \ref{cmass_flux_limit}). The horizontal line indicates the CMASS
  $d_{\rm perp}$ cut (equation \ref{cmass_dperp}). Tilted black lines
  indicate the fiducial sliding cut (equation \ref{cmass_sliding}) as
  well as the sliding cut that defines the sparsely sampled region
  (equation \ref{cmass_sliding_sparse}). The vertical error bar
  indicates the typical scatter between the target photometry and the
  co-add photometry. The horizontal error bar indicates the typical
  scatter for $r_{\rm cmod}$. The stellar mass completeness of CMASS
  is due to the fact that the intrinsic color distributions of
  galaxies at fixed mass often extend beyond the BOSS color
  boundaries, as well as to scatter across these color boundaries
  because of the shallower depth of the target photometry. The later
  effect is not present in this Figure which is based on co-add
  photometry.}
\label{cmass_birth_parameter}
\end{center}
\end{figure*}

Figure \ref{cmass_birth_parameter} shows galaxies from the {\sc
  s82-mgc} in the CMASS redshift range as a function of $i_{\rm cmod}$
and $d_{\rm perp}$. Because a number of galaxies in Figure
\ref{cmass_birth_parameter} are not contained in the original BOSS
target catalog\footnote{bosstarget-lrg-main007-collate.fits. See
  Appendix.}, $i_{\rm cmod}$ and $d_{\rm perp}$ are taken from the
deeper Stripe 82 co-add photometry \citep[][]{Annis:2011}. For CMASS
galaxies, the typical RMS scatter between the co-add photometry and
the shallower photometry of the target catalog is of order
$\sigma_{\rm RMS}=0.08$ mag for $d_{\rm perp}$ and
$\sigma_{\rm RMS}=0.14$ mag for $i_{\rm cmod}$.

Figure \ref{cmass_birth_parameter} clearly demonstrates that the CMASS
sample is only complete in terms of mass and color at the highest
masses ($\log_{10}(M_*/M_{\odot})>11.7$) and in a narrow redshift
window at $z\sim0.6$. The $d_{\rm perp}$ cut mainly affects the sample
selection at lower redshifts ($z< 0.6$). The flux limit affects the
sample selection at low stellar masses ($\log_{10}(M_*/M_{\odot})<
11.5$) and at higher redshifts ($z>0.6$) but is relatively unimportant
for massive galaxies at low redshifts. Galaxies with high values of
$b_{1000}$ (suggesting recent star-formation) are excluded from the
sample at low redshift due to a combination of all three cuts. At
higher redshifts ($z>0.6$), the sample is mainly flux limited and
includes a larger range of galaxy colors at fixed magnitude.

% /Users/alexie/Work/Boss/Completeness/Pycode/cmass_colors.py

%>> Cpara :
%Mean :    0.0364601
%sigma :    0.0658348
%>> Rmodel :
%Mean :   -0.0262746
%sigma :     0.110495

\begin{figure*}
\begin{center}
\includegraphics[width=14cm]{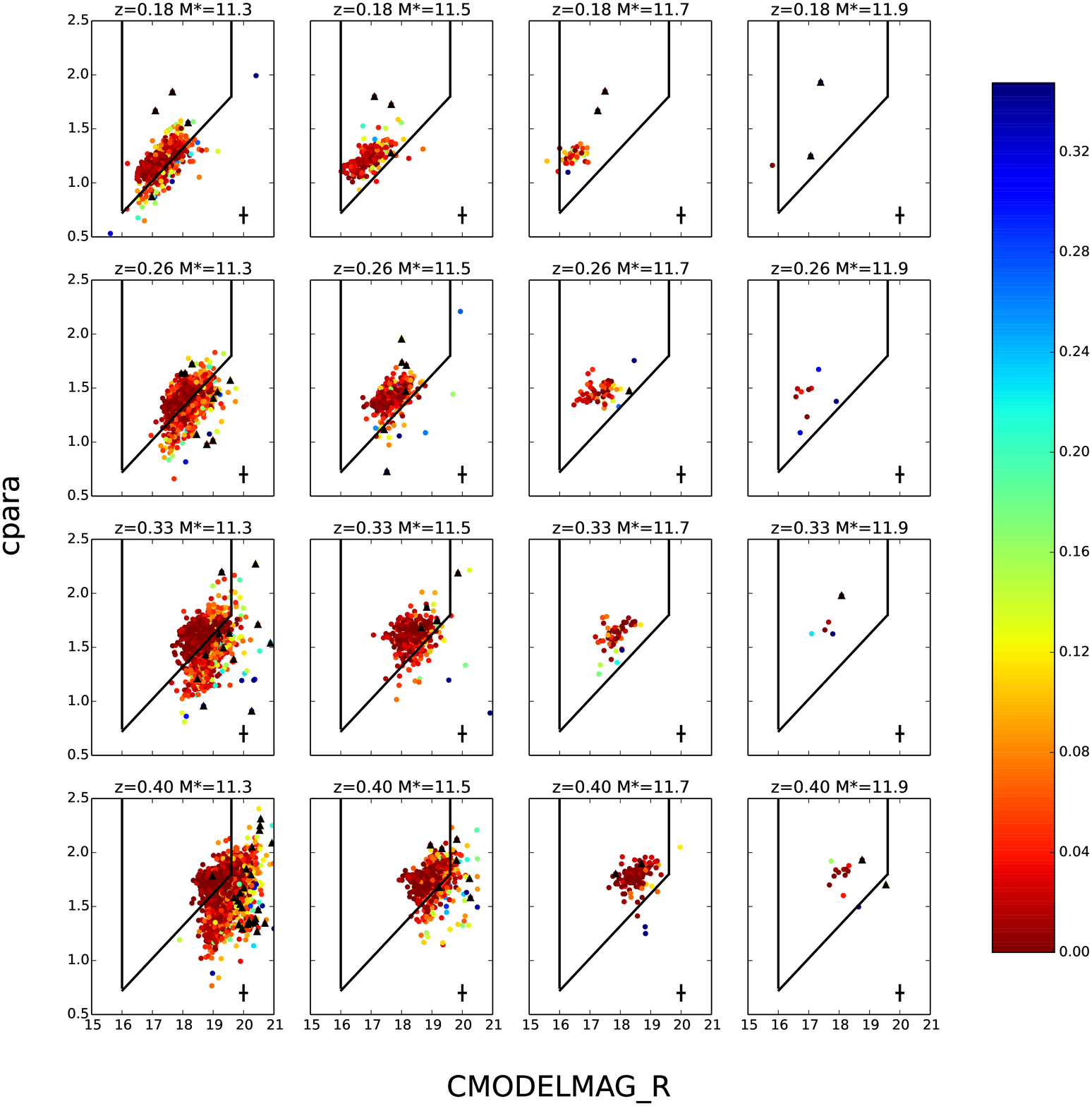}
\caption{Galaxies from the {\sc s82-mgc} in the LOWZ redshift range as
  a function of $r_{\rm cmod}$ and $c_\parallel$. Galaxies are color
  coded according to their birth parameter $b_{1000}$: larger values
  of $b_{1000}$ (bluer colors) indicate galaxies that have experienced
  larger amounts of star formation in the past Gyr. Vertical black
  lines indicate the LOWZ flux limits (equation
  \ref{lowz_flux_limit}). The tilted black line represents the LOWZ
  sliding cut (equation \ref{lowz_sliding}). Black triangles indicate
  galaxies that are removed by the $c_\perp$ cut (equation
  \ref{lowz_cperp}). The vertical error bar indicates the typical
  scatter between the target photometry and the co-add photometry. The
  horizontal error bar indicates the typical scatter for
  $r_{\rm cmod}$.}
\label{lowz_birth_parameter}
\end{center}
\end{figure*}

Figure \ref{lowz_birth_parameter} presents $r_{\rm cmod}$ and
$c_{\parallel}$ in the LOWZ redshift range where $r_{\rm cmod}$ and
$c_{\parallel}$ are taken from the {\sc s82-mgc}. For LOWZ galaxies,
the typical RMS scatter between the co-add photometry and the
shallower photometry of the target catalog is of order
$\sigma_{\rm RMS}=0.07$ mag for $c_{\parallel}$ and
$\sigma_{\rm RMS}=0.11$ mag for $r_{\rm cmod}$. Figure
\ref{lowz_birth_parameter} shows that the LOWZ sample is roughly
complete in terms of mass and color at $\log_{10}(M_*/M_{\odot})>11.6$
over the redshift range $0.15<z<0.43$. At lower stellar masses, the
mass completeness is limited by the $c_{\parallel}$ sliding cut. The
flux limit mainly affects the sample selection at low stellar masses
($\log_{10}(M_*/M_{\odot})< 11.7$) and at higher redshifts
($z > 0.3$).

% /Users/alexie/Work/Boss/Completeness/Pycode/lowz_colors.py

%%%%%%%%%%%%%%%%%%%%%%%%%%%%%%%%%%%%%%%%%%%%%%%%%%%%%%%%%%%%%%%%%%%%%%%%%%%%%%
%     CMASS COMP
%%%%%%%%%%%%%%%%%%%%%%%%%%%%%%%%%%%%%%%%%%%%%%%%%%%%%%%%%%%%%%%%%%%%%%%%%%%%%%

\section{Total Stellar Mass Functions}\label{totalsmf}

To estimate the completeness of the BOSS samples, we first estimate
the total stellar mass function in four redshift bins. For a given
mass bin and redshift, the stellar mass completeness of any given
spectroscopic sample, $A$, will be estimated via the ratio
$c=\phi_{\rm A}/\phi_{\rm tot}$ where $\phi_{\rm A}$ is the number
density of sample A and $\phi_{\rm tot}$ is the number density of the
total stellar mass function.

We construct redshift bins that correspond to a volume of 0.05--0.06
Gpc$^3$.  For the LOWZ sample, this requires a single large redshift
bin, $z_1=[0.15,0.43]$. For CMASS, we divide the sample into three
equal volume bins: $z_2=[0.43,0.54]$, $z_3=[0.54,0.63]$, and
$z_4=[0.63,0.7]$.  The time span for these redshift bins is 2.6 Gyr,
0.8 Gyr, 0.6 Gyr, and 0.4 Gyr, respectively. We take the
redshift-binned mass functions to represent the galaxy distribution
sampled at the midpoint of each bin.  A more detailed investigation of
the redshift evolution of the total SMF is presented and discussed in
Paper III (Bundy et al. in prep).

To construct the total SMF, we use $z_{\rm best}$ and $M_{\rm best}$
from the {\sc s82-mgc} to measure the SMF at the high-mass end and
data from PRIMUS \citep[][]{Moustakas:2013} to evaluate the low-mass
end.  Because we use a parent sample that is complete above
$\log M_*/M_{\odot} = 11.2$ for $z < 0.7$, $V_{\rm max}$ corrections
are not required.  Errors on the Stripe 82 SMFs are derived via
bootstrap using 214 roughly equal-area bootstrap regions.

\begin{figure}
\begin{center}
\includegraphics[width=8.8cm]{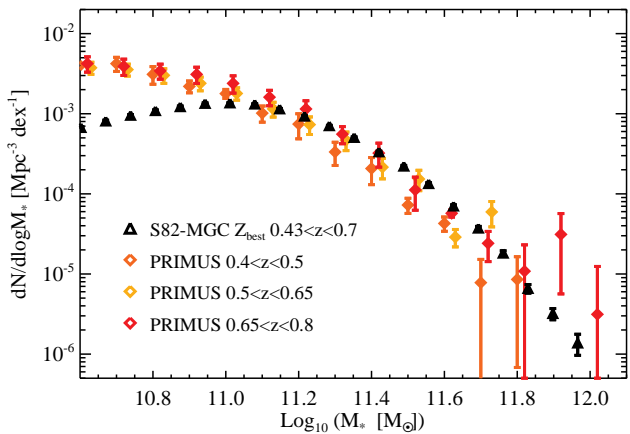}
\caption{Total stellar mass function derived using $z_{\rm best}$ for
  $0.43<z<0.7$. Diamonds show the stellar mass functions from PRIMUS
  in three redshift bins from $z=0.4$ to $z=0.8$. Our Stripe 82 sample
  tightly constrains the high-mass end of the stellar mass function
  ($\log_{10}(M_*/M_{\odot})>11.5$) while PRIMUS constrains the low
  mass end. The Stripe 82 sample is mass-complete to
  $\log_{10}(M_*/M_{\odot})=11.2$ at $z=0.7$.}
\label{combined_smf}
\end{center}
\end{figure}

Figure \ref{combined_smf} shows mass functions from PRIMUS and from
the {\sc s82-mgc} over $0.43<z<0.7$.  The figure demonstrates two key
points. First, it is clear that mass functions from PRIMUS and COSMOS
are insufficient to constrain the SMF at masses above
$\log_{10}(M_*/M_{\odot})=11.5$ --- emphasizing the importance of the
{\sc s82-mgc} at these high masses. Second, as expected, the SMF of
the {\sc s82-mgc} UKWIDE sample itself begins to turn over due to
incompleteness around $\log_{10}(M_*/M_{\odot})\sim11.2$ at
$z\sim0.7$.

\begin{figure*}
\begin{center}
\includegraphics[width=16cm]{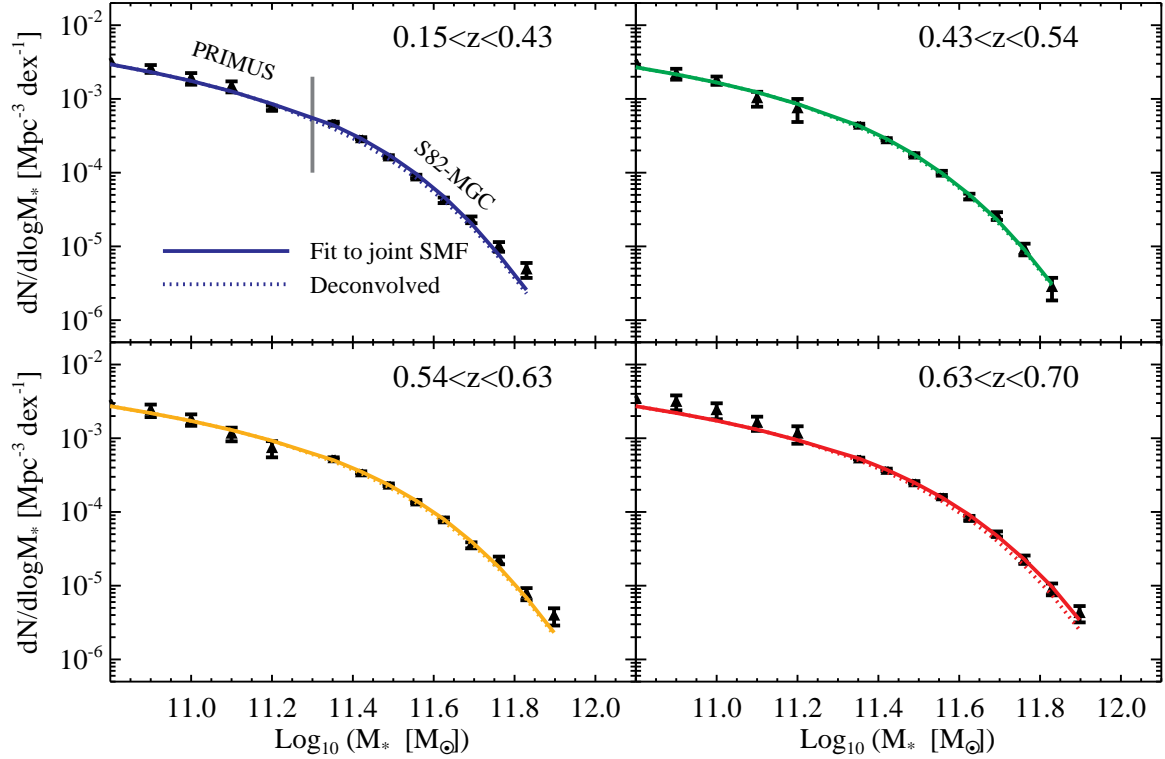}
\caption{Total SMF in four redshift bins from $z=0.15$ to $z=0.7$. To
  construct the total SMF, we combine data from the PRIMUS survey at
  $\log_{10}(M_*/M_{\odot})<11.3$ with data from the {\sc s82-mgc} at
  $\log_{10}(M_*/M_{\odot})>11.3$. Solid lines indicate our fit to the
  total SMF and dotted lines show the total SMF deconvolved for the
  effects of scatter introduced by photometric redshift. Because a
  large fraction of galaxies at the high-mass end have a spectroscopic
  redshift ($>$80\% at the high mass end), the inclusion of galaxies
  with photometric redshifts has a negligible impact on the SMF.}
\label{smf_funct_z_4panels}
\end{center}
\end{figure*}

To obtain total SMFs for our subsequent analysis, we combine SMF
measurements from the PRIMUS survey at $\log_{10}(M_*/M_{\odot})<11.3$
with those from the {\sc s82-mgc} at
$\log_{10}(M_*/M_{\odot})>11.3$. For PRIMUS we use the mass functions
estimated by \citet[][]{Moustakas:2013} in the redshift range closest
to our bin\footnote{Small offsets ($\sim$0.1 dex) are likely between
  the PRIMUS stellar masses and the masses from the {\sc
    s82-mgc}. However, PRIMUS did not compute stellar masses for their
  Stripe 82 field due to a lack of Spitzer data. Hence, we cannot
  directly compare mass estimates between PRIMUS and the {\sc
    s82-mgc}. However, we expect that any offsets will only have a
  minor impact on completeness estimates and only for
  $\log_{10}(M_*/M_{\odot})<11.3$.}. The PRIMUS mass functions assume
a Chabrier IMF, BC03 stellar population models, and have been adjusted
to match our fiducial cosmology. Figure \ref{smf_funct_z_4panels}
presents the total SMF in our four redshift bins from $z=0.15$ to
$z=0.7$.

The {\sc s82-mgc} $z_{\rm best}$ parameter uses a combination of
spectroscopic redshifts and photometric redshifts. Because redshift
errors translate into an error on the luminosity distance, galaxies
that do not have a spectroscopic redshift will have an extra error
term for $M_{\rm best}$ which we denote as $\sigma_{\rm M*-zp}$. In
the {\sc s82-mgc}, this error term ranges from about
$\sigma_{\rm M*-zp}=0.05$ dex to $\sigma_{\rm M*-zp}=0.1$ dex. One
concern with our approach is that the inclusion of galaxies with
photometric redshifts will cause an increase (due to Eddington bias)
in the amplitude of the steep, high-mass end of the total SMF compared
to a scenario in which all galaxies in our sample had a spectroscopic
redshift. This behavior would lead our completeness estimates to be
underestimated because $\phi_{\rm tot}$ would be artificially inflated
by this additional scatter term relative to the number densities
obtained for a spectroscopic sample.

We account for this effect by forward modeling the SMF by convolving
for $M_*$ scatter induced by photometric redshift uncertainties.  We
assume an input functional form for the total SMF that follows a
double Schechter function \citep{Baldry:2008}:

\begin{displaymath}
\phi(M_{*}) =  (\ln 10)\exp\left[-\frac{M_{*}}{M_{0}}\right] \times
\hspace{0.65\columnwidth}
\end{displaymath}
\vspace{-3ex}
\begin{equation}
\quad
\left\{ \phi_{1}10^{(\alpha_{1}+1)(\log M_{*} -\log M_{0})}
           + \phi_{2}10^{(\alpha_{2}+1)(\log M_{*} -\log M_{0})} \right\}
\end{equation}

%/Users/alexie/Work/Boss/Completeness/Pros/total_cmass_smf_schecter_params.txt
\noindent where $\alpha_{2}>\alpha_{1}$ and the second term dominates
at the low-mass end. We generate Monte Carlo realizations of this
function that sample various ranges for its parameters.  The values of
$\phi_{1}$, $\phi_{2}$ and $M_{0}$ are allowed to vary, while we fix
$\alpha_{1}=-0.46$ and $\alpha_{2}=-1.58$.  A mock sample is drawn
from each realization of the input SMF and additional scatter,
$\sigma_{\rm M*-zp}$, is added to the mock stellar masses of this
sample following the estimated $\sigma_{\rm M*-zp}(M_*,z)$
distribution from the {\sc s82-mgc}. In our Monte
Carlo mock realizations, the fraction of galaxies that have a
spectroscopic redshift (i.e, $\sigma_{\rm M*-zp}=0$) as a function of
mass and redshift is identical to the {\sc s82-mgc}.

The results of fitting our mock samples to the observed SMF data are
presented in Figure \ref{smf_funct_z_4panels}. Because a large
fraction of galaxies at the high-mass end have a spectroscopic
redshift, the inclusion of galaxies with photometric redshifts only
has a very minor impact on the SMF.  The effects of
$\sigma_{\rm M*-zp}$ are hence negligible for the {\sc s82-mgc}. The
best fit values for the double Schecter fits are given in Table
\ref{totalsmf_params}.  In this exercise, we have not accounted for
additional sources of scatter in $M_*$, namely from the mass estimates
themselves or from uncertainties in total luminosities.  Because these
latter sources of errors are also present in the spectroscopic
samples, they cancel in our completeness functions which divide spec-z
SMFs by those supplemented with photo-zs.

% /Users/alexie/Work/Boss/Completeness/TotalSMF/total_smf_schecter_params_0.15_0.43.txt
% => copied final versions to /Users/alexie/Work/Papers/Boss_completeness/Files_used_in_paper
\begin{table*}
  \caption{Parameters of the double Schecter fit to the total SMF as a
    function of redshift. Only the first three parameters are varied
    in the fit. These parameters correspond to the SMF after
    deconvolving for the effects of scatter due to the inclusion of a
    sub-set of galaxies with photometric redshifts.}\label{totalsmf_params}
\begin{tabular}{@{}lccccc}
\hline
Redshift &$\log_{10}(\phi_{1}/$Mpc$^{-3}$dex$^{-1})$ & $\log_{10}(\phi_{2}/$Mpc$^{-3} $dex$^{-1})$ & $\log_{10}(M_0/M_{\odot})$&$\alpha_{1}$  &$\alpha_{2}$ \\
\hline
$z_1=[0.15,0.43]$ &  $-2.97$     &   $-2.79$    &  $10.910$ &  $-0.46$& $-1.58$ \\
$z_2=[0.43,0.54]$ &  $-2.95$     &   $-2.89$     &  $10.922$  & $-0.46$ &$-1.58$ \\
$z_3=[0.54,0.63]$ &   $-3.06$    &   $-2.91$    &  $10.986$  & $-0.46$ & $-1.58$\\
$z_4=[0.63,0.7]$ &   $-3.06$   &    $-2.92$    &   $10.995$  &  $-0.46$&$-1.58$\\
\hline
\end{tabular}
\end{table*}

%%%%%%%%%%%%%%%%%%%%%%%%%%%%%%%%%%%%%%%%%%%%%%%%%%%%%%%%%%%%%%%%%%%%%%%%%%%%%%
%     COMPLETENESS
%%%%%%%%%%%%%%%%%%%%%%%%%%%%%%%%%%%%%%%%%%%%%%%%%%%%%%%%%%%%%%%%%%%%%%%%%%%%%%

\section{Stellar Mass Completeness of BOSS samples}\label{comp}

%mean redshift CMASS :      0.549080
%mean redshift LOWZ :      0.287720

\subsection{Stellar Mass Completeness of CMASS}\label{sect_cmass_comp}

\begin{figure*}
\begin{center}
\includegraphics[width=17cm]{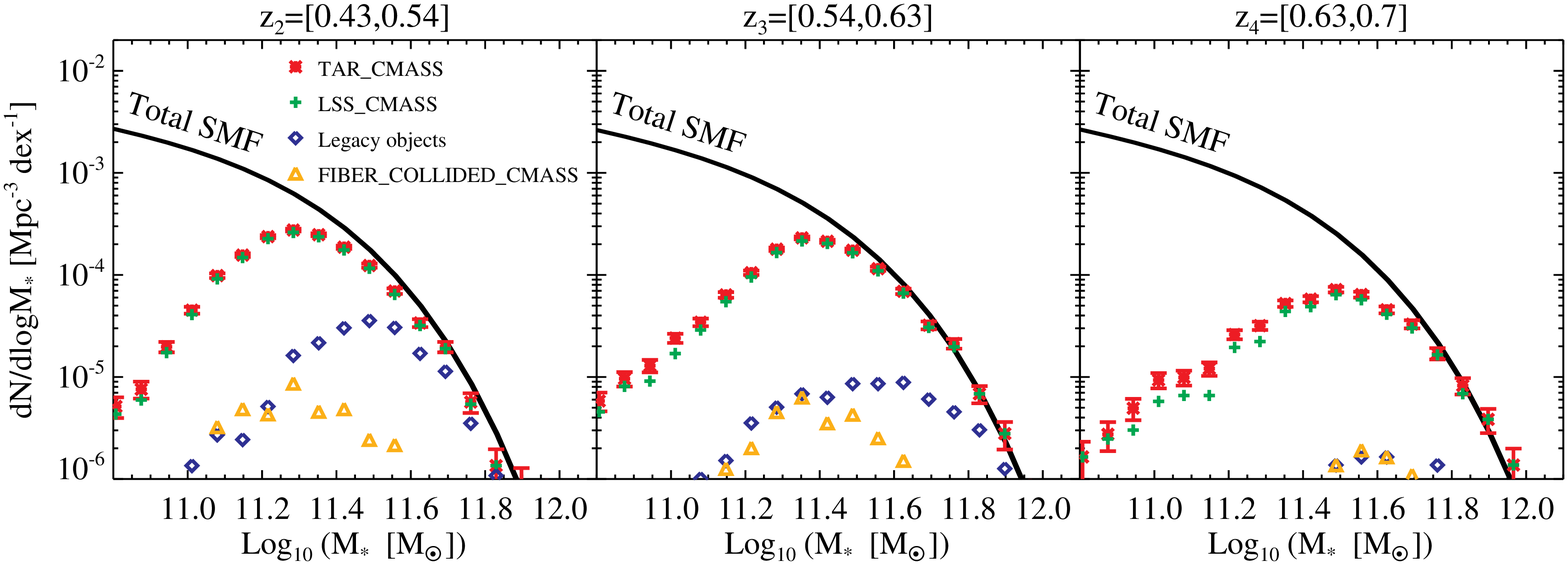}
\caption{Total SMF compared to target CMASS galaxies (red stars),
  fiber collided galaxies (orange triangles), CMASS galaxies from the
  large-scale structure catalog (green plus symbols), and Legacy
  objects (blue diamonds) in three different redshift bins. We
  sub-divide each of these redshift bins into two roughly equal volume
  bins to compute the completeness of CMASS in a total of six redshift
  bins. At low redshifts, legacy objects make up $50\%$ of galaxies at
  the high mass end.}
\label{cmass_types}
\end{center}
\end{figure*}

With the total SMF now in hand, we derive the stellar mass
completeness for the LSS\_CMASS sample. Completeness is estimated by
comparing the total number of LSS\_CMASS galaxies in a given redshift
and stellar mass bin to that derived from the total SMF. For the total
SMF, we use the three redshift bins $z_2$, $z_3$, and $z_4$. Figure
\ref{cmass_types} presents our total SMF compared to target CMASS
galaxies, fiber collided galaxies, CMASS galaxies from the LSS
catalog, and Legacy objects, in our three fiducial redshift bins.

\begin{figure*}
\begin{center}
\includegraphics[width=16cm]{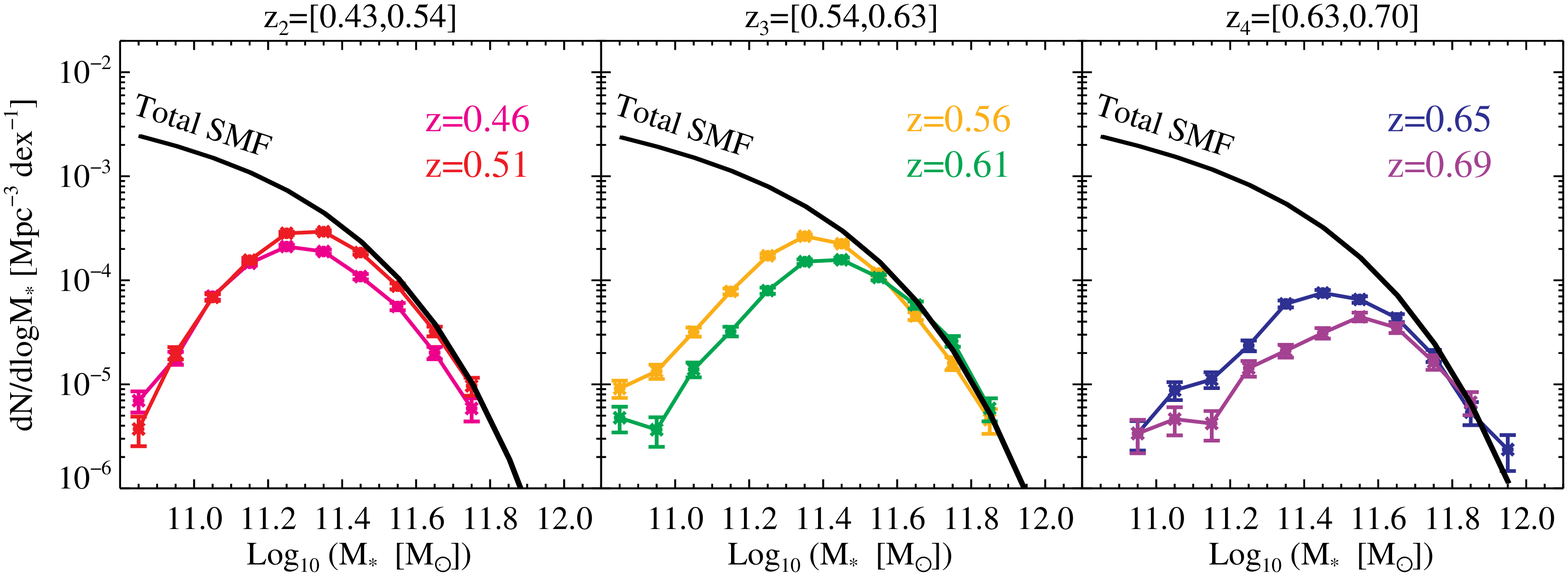}
\caption{Stellar mass functions of the LSS\_CMASS sample in six
  redshift bins compared to the total SMF. Completeness is estimated
  via the ratio $c=\phi_{\rm A}/\phi_{\rm tot}$ where $\phi_{\rm A}$
  is the number density of CMASS in a given redshift bin and
  $\phi_{\rm tot}$ is the number density of the total stellar mass
  function.}
\label{cmass_smfs_z}
\end{center}
\end{figure*}

\begin{figure*}
\begin{center}
\includegraphics[width=16cm]{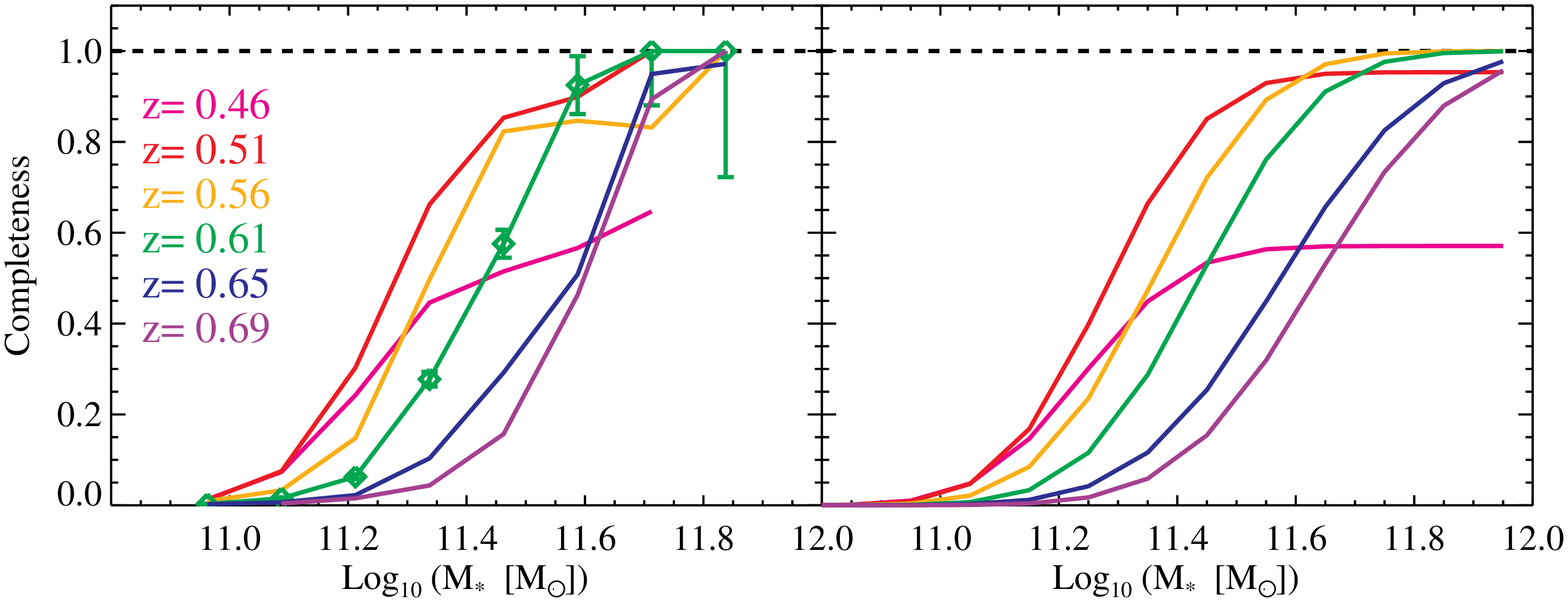}
\caption{Left: stellar mass completeness of the CMASS sample as a
  function of stellar mass and redshift. For clarity, errors are shown
  only for the $z=0.61$ redshift bin. Right: fits to the completeness
  using Equation \ref{ncen}.}
\label{cmass_comp}
\end{center}
\end{figure*}

We further sub-divide each of these redshift bins into two roughly
equal volume bins to compute the completeness of CMASS in a total of
six redshift bins. The CMASS SMFs for each of these six redshift bins
are shown in Figure \ref{cmass_smfs_z}. If the number of LSS\_CMASS
galaxies fluctuates above the prediction based on the total SMF, the
completeness is simply set to unity. The results are presented in
Figure \ref{cmass_comp} and the completeness values are given in Table
\ref{cmass_comp_table}.

For convenience, we also fit the completeness with the following
functional form:

\begin{equation}\label{ncen}
c = \frac{f}{2}\left[
  1+\mbox{erf}\left(\frac{\log_{10}( M_{*}/M_1)}{\sigma}\right)\right],
\end{equation}

\noindent with free parameters $f$, $\sigma$, and $M_1$. The results
are shown on the right hand side of Figure \ref{cmass_comp} and the values of
the fitted parameters are given in Table \ref{cmass_comp_table2}.

%/Users/alexie/Work/Papers/Boss_completeness/Files_used_in_paper/cmass_measure_completeness.txt
\begin{table*}
\caption{Measured stellar mass completeness of the CMASS sample. Bins
  with less than 10 galaxies are marked with an 'x' symbol. The bottom
line corresponds to the completeness estimates for the combined
LOW+CMASS sample at $0.38<z<0.48$.}\label{cmass_comp_table}
\begin{tabular}{@{}lccccccc}
\hline
z&$\log_{10}(M_*/M_{\odot})$&$\log_{10}(M_*/M_{\odot})$&$\log_{10}(M_*/M_{\odot})$&$\log_{10}(M_*/M_{\odot})$&$\log_{10}(M_*/M_{\odot})$&$\log_{10}(M_*/M_{\odot})$&$\log_{10}(M_*/M_{\odot})$\\
 &$=$11.09 &$=$11.21&$=$11.34&$=$11.46&$=$11.59&$=$11.71&$=$11.84\\
\hline
$0.46$&$0.07\pm0.01$&$0.24\pm0.01$&$0.45\pm0.03$&$0.51\pm0.04$&$0.57\pm0.06$&$0.65\pm0.11$&x\\
$0.51$&$0.08\pm0.00$&$0.30\pm0.01$&$0.66\pm0.03$&$0.85\pm0.05$&$0.90\pm0.07$&$1.00\pm0.17$&x\\
$0.56$&$0.03\pm0.00$&$0.15\pm0.01$&$0.50\pm0.02$&$0.82\pm0.04$&$0.85\pm0.06$&$0.83\pm0.10$&$1.00\pm0.23$\\
$0.61$&$0.02\pm0.00$&$0.06\pm0.01$&$0.28\pm0.02$&$0.58\pm0.03$&$0.92\pm0.06$&$1.00\pm0.12$&$1.00\pm0.28$\\
$0.65$&$0.01\pm0.00$&$0.02\pm0.00$&$0.10\pm0.01$&$0.29\pm0.02$&$0.51\pm0.04$&$0.95\pm0.09$&$0.97\pm0.24$\\
$0.69$&$0.00\pm0.00$&$0.02\pm0.00$&$0.04\pm0.01$&$0.16\pm0.01$&$0.46\pm0.05$&$0.89\pm0.11$&$1.00\pm0.39$\\
\hline
comb&$0.05\pm0.00$&$0.16\pm0.01$&$0.42\pm0.02$&$0.68\pm0.03$&$0.92\pm0.06$&$1.00\pm0.13$&$1.00\pm0.41$\\
\hline
\end{tabular}
\end{table*}

%/Users/alexie/Work/Papers/Boss_completeness/Files_used_in_paper/cmass_measure_completeness_ncen.txt
\begin{table}
  \caption{Functional fits for completeness of the LSS\_CMASS
    sample. The bottom line corresponds to the combined LOWZ+CMASS sample.}\label{cmass_comp_table2}
\begin{tabular}{@{}lccc}
\hline
Redshift & $f$ & $\sigma$ & $\log_{10}(M_{1}/M_{\odot})$ \\
\hline
$0.46$&$0.57$&$0.20$&$11.24$\\
$0.51$&$0.95$&$0.20$&$11.28$\\
$0.56$&$1.00$&$0.22$&$11.36$\\
$0.61$&$1.00$&$0.22$&$11.44$\\
$0.65$&$1.00$&$0.27$&$11.57$\\
$0.69$&$1.00$&$0.26$&$11.64$\\
\hline
comb&$1.00$&$0.25$&$11.38$\\
\hline
\end{tabular}
\end{table}

Figure \ref{cmass_comp} demonstrates that CMASS is 80\% complete at
$\log_{10}(M_*/M_{\odot}) > 11.6$ in the narrow redshift range
$z=[0.51,0.61]$. At the mean redshift of the CMASS sample,
$\overline{z}=0.55$, CMASS is roughly 80\% complete at
$\log_{10}(M_*/M_{\odot})=11.4$. For comparison,
\citet[][]{Maraston:2013} conclude that BOSS is complete above
$\log_{10}(M_*/M_{\odot})=11.3$ at $z<0.6$ and at
$\log_{10}(M_*/M_{\odot})=11.6$ at $z>0.6$. This work narrows this
statement to $z=[0.51,0.61]$ and shows that the completeness decreases
at lower and higher redshifts. Based on these considerations,
referring to this sample in terms of ``constant mass'' should only be
considered in loose terms.

%%%%%%%%%%%%%%%%%%%%%%%%%%%%%%%%%%%%%%%%%%%%%%%%%%%%%%%%%%%%%%%%%%%%%%%%%%%%%%
%     LOWZ COMP
%%%%%%%%%%%%%%%%%%%%%%%%%%%%%%%%%%%%%%%%%%%%%%%%%%%%%%%%%%%%%%%%%%%%%%%%%%%%%%
%
\subsection{Stellar Mass Completeness of LOWZ}\label{sect_lowz_comp}
\begin{figure}
\begin{center}
\includegraphics[width=8.8cm]{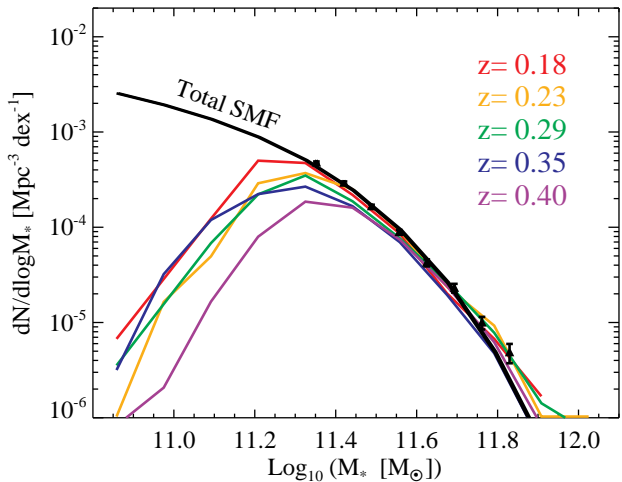}
\caption{Stellar mass functions of the LSS\_LOWZ sample in five
  redshift bins compared to the total SMF. Completeness is estimated
  via the ratio $c=\phi_{\rm A}/\phi_{\rm tot}$, where $\phi_{\rm A}$
  is the number density of LOWZ in a given redshift bin and
  $\phi_{\rm tot}$ is the number density of the total stellar mass
  function. In this redshift bin, our double Schecter fit slightly
  under-estimates the total SMF at the high-mass end at
  $\log_{10}(M_*/M_{\odot})\sim11.8.$}
\label{lowz_smfs}
\end{center}
\end{figure}

\begin{figure*}
\begin{center}
\includegraphics[width=16cm]{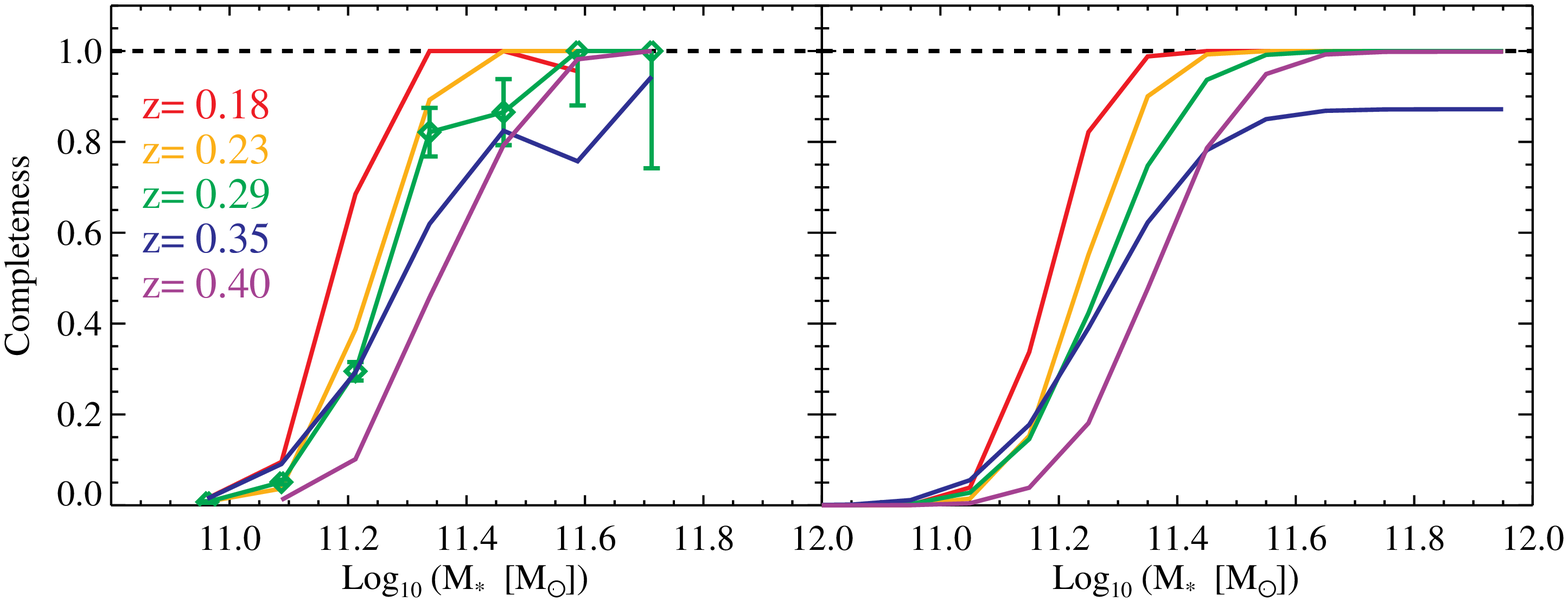}
\caption{Left: stellar mass completeness of the LOWZ sample as a
  function of stellar mass and redshift. For clarity, errors are shown
  only for the $z=0.29$ redshift bin. Right: fits to the completeness
  using Equation \ref{ncen}.}
\label{lowz_comp}
\end{center}
\end{figure*}

We now proceed in a similar manner for the LOWZ sample in the redshift
range $0.15<z<0.43$. For LOWZ, we use the total SMF estimated in the
the $z_1=[0.15,0.43]$ redshift bin and shown in Figure
\ref{smf_funct_z_4panels}. Figure \ref{lowz_smfs} presents the SMFs of
the LOWZ sample compared to the total SMF as a function of
redshift. In a similar fashion as in the previous section,
completeness is estimated by comparing the total number of LSS\_LOWZ
galaxies in a given redshift and stellar mass bin compared to the
expectation derived from the total SMF. The results are displayed in
Figure \ref{lowz_comp} and the completeness values are listed in Table
\ref{lowz_comp_table}. The values of the functional fits to the
completeness are given in Table \ref{lowz_comp_table2}.

%/Users/alexie/Work/Papers/Boss_completeness/Files_used_in_paper/lowz_measure_completeness.txt
\begin{table*}
\caption{Measured stellar mass completeness of the LOWZ sample. }\label{lowz_comp_table}
\begin{tabular}{@{}lccccccc}
\hline
z&$\log_{10}(M_*/M_{\odot})$&$\log_{10}(M_*/M_{\odot})$&$\log_{10}(M_*/M_{\odot})$&$\log_{10}(M_*/M_{\odot})$&$\log_{10}(M_*/M_{\odot})$&$\log_{10}(M_*/M_{\odot})$\\
 &$=$11.09 &$=$11.21&$=$11.34&$=$11.46&$=$11.59&$=$11.71\\
\hline
$0.18$&$0.10\pm0.01$&$0.69\pm0.04$&$1.00\pm0.07$&$1.00\pm0.10$&$0.96\pm0.15$&x\\
$0.23$&$0.04\pm0.01$&$0.39\pm0.03$&$0.89\pm0.07$&$1.00\pm0.10$&$1.00\pm0.16$&$1.00\pm0.35$\\
$0.29$&$0.05\pm0.01$&$0.29\pm0.02$&$0.82\pm0.05$&$0.87\pm0.07$&$1.00\pm0.12$&$1.00\pm0.26$\\
$0.35$&$0.09\pm0.01$&$0.29\pm0.02$&$0.62\pm0.04$&$0.82\pm0.06$&$0.76\pm0.09$&$0.94\pm0.18$\\
$0.40$&$0.01\pm0.00$&$0.10\pm0.01$&$0.46\pm0.03$&$0.79\pm0.05$&$0.98\pm0.10$&$1.00\pm0.19$\\
\hline
\end{tabular}
\end{table*}

%/Users/alexie/Work/Papers/Boss_completeness/Files_used_in_paper/lowz_measure_completeness_ncen.txt
\begin{table}
\caption{Functional fits for completeness of the LSS\_LOWZ sample.}\label{lowz_comp_table2}
\begin{tabular}{@{}lccc}
\hline
Redshift & $f$ & $\sigma$ & $\log_{10}(M_{1}/M_{\odot})$ \\
\hline
$0.18$&$1.00$&$0.11$&$11.18$\\
$0.23$&$1.00$&$0.12$&$11.24$\\
$0.29$&$1.00$&$0.16$&$11.27$\\
$0.35$&$0.87$&$0.20$&$11.27$\\
$0.40$&$1.00$&$0.17$&$11.36$\\
\hline
\end{tabular}
\end{table}

The errors on the LOW completeness are large at the high mass end
because the sub-volumes used to measure the redshift dependance of the
completeness are relatively small. As a result, for example, the
best-fitting functional form does not converge to unity at large
masses in the $z=0.35$ bin but does in the $z=0.29$ and $0.40$
redshift bins. The values provided in this paper should be considered
as {\em estimates} with relatively large errors at the high mass
end. Studies that are sensitive to these completeness estimates must
take the reported errors in Table \ref{lowz_comp_table} into
consideration.

Figure \ref{lowz_comp} demonstrates that LOWZ is at least 80\%
complete at $\log_{10}(M_*/M_{\odot}) > 11.6$ over the entire
redshift range and at least 90\% complete at
$\log_{10}(M_*/M_{\odot}) > 11.5$ in the redshift range
$z=[0.18,0.29]$. At the mean redshift of the LOWZ sample,
$\overline{z}=0.29$, LOWZ is roughly 80\% complete at
$\log_{10}(M_*/M_{\odot})=11.4$. Interestingly, these results counter
the conventional wisdom that CMASS is more complete at higher stellar
masses compared to LOWZ \citep[\eg][]{Anderson:2014}.

Finally, although we do find a high stellar mass completeness for
LOWZ, \citet{Hoshino:2015} reported that the $r_{\rm cmod}>16$ cut in
equation \ref{lowz_flux_limit} removes a small fraction ($\sim$5\%) of
Brightest Cluster Galaxies from the nominal LOWZ sample.

\subsection{Combined Sample}\label{sect_combined_comp}

\begin{figure}
\begin{center}
\includegraphics[width=8.5cm]{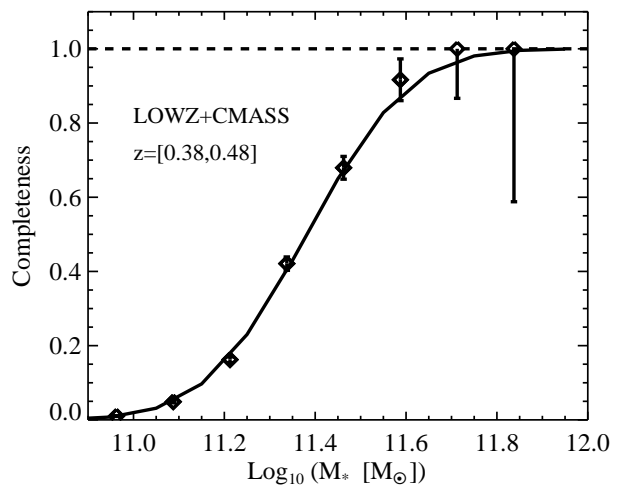}
\caption{Stellar mass completeness of the combined LOWZ+CMASS sample
  in the redshift range $0.38<z<0.48$. Legacy objects are included in
  this combined sample. The solid line corresponds to the best-fit
  using Equation \ref{ncen}. The combined sample is roughly 80\%
  complete to $\log_{10}(M_*/M_{\odot})=11.6$.}
\label{comb_comp}
\end{center}
\end{figure}

As can be seen from Figure \ref{nden_fig}, CMASS and LOWZ overlap
in the redshift range $z\sim0.38$ to $z\sim0.48$. At these redshifts,
it may be useful for certain studies to combine the two samples
together. The stellar mass completeness of the combined LOWZ and CMASS
samples in the redshift range $0.38<z<0.48$ are presented in Figure
\ref{comb_comp}. Legacy objects are also included in this combined
sample. The combined sample is roughly 80\% complete to
$\log_{10}(M_*/M_{\odot})=11.6$. The completeness values for the
combined samples are appended to the bottom of Table
\ref{cmass_comp_table} and Table \ref{cmass_comp_table2}.

In conclusion, the {\em combination} of LOW and CMASS (and Legacy objects)
is 80\% complete to $\log_{10}(M_*/M_{\odot})=11.6$ at $z<0.61$.

%%%%%%%%%%%%%%%%%%%%%%%%%%%%%%%%%%%%%%%%%%%%%%%%%%%%%%%%%%%%%%%%%%%%%%%%%%%%%%
%     PREVIOUS WORK
%%%%%%%%%%%%%%%%%%%%%%%%%%%%%%%%%%%%%%%%%%%%%%%%%%%%%%%%%%%%%%%%%%%%%%%%%%%%%%

\section{Samples from Previous Studies using BOSS Data}\label{previous_work}

In this section, we evaluate the potential impact of stellar mass
incompleteness for specific samples that have been used in previous
BOSS analyses.  We focus on the samples used in \citet{Maraston:2013},
\citet{Shankar:2014}, \citet{Miyatake:2013}, and \citet{More:2014}.

\subsection{CMASS Mass Functions from \citet{Maraston:2013}}\label{m13smf}

\citet{Maraston:2013} computed stellar masses for BOSS galaxies using
SED fits to the SDSS single epoch {\it ugriz} photometry. SED fits
were performed using the HYERSPECZ code \citep[][]{Bolzonella:2000}
using the \citet[][]{Maraston:2011} stellar population libraries and
assuming a Kroupa IMF \citep[][]{Kroupa:2001}. Masses were computed
for all BOSS galaxies in two separate runs: one using a passive
template and one using a suite of star-forming templates with
exponentially declining or truncated star-formation histories. The
stellar masses of star-forming galaxies were adjusted upwards by 0.25
dex to account for a bias when fitting star forming galaxies in which
the best-fit model may underestimate the total stellar mass because of
fitting the brightest population \citep[][]{Maraston:2010}. This bias
was identified using mock catalogs. For both templates, masses with
and without mass loss were computed in order to allow comparisons with
the literature. These stellar masses are a standard output of the BOSS
pipeline and are publicly released.

The two separate runs were combined into a single stellar mass catalog
by adopting the star-forming templates for BOSS galaxies with apparent
$g-i$ colors less than 2.35 and the passive template for galaxies with
$g-i>2.35$. This matched the empirical morphological mix determined
from the COSMOS field by \citet[][]{Masters:2011}

\citet{Maraston:2013} computed the mass function for the CMASS sample
in three redshift bins from $z=0.45$ to $z=0.7$ with the aim at
constraining the assembly of the most massive galaxies in relation to
galaxy formation models. Because the stellar mass completeness of the
CMASS sample was not known at that time, \citet{Maraston:2013} did not
apply any completeness corrections to these CMASS SMF, opting instead
to apply the BOSS selection cuts to semi-analytic models (SAMs) when
comparing with the theoretical mass function from galaxy formation
models. \citet{Maraston:2013} find a deficit of massive galaxies
($\log(M_*/M_{\odot})>11.3$~ for a Kroupa IMF) between the BOSS data
and SAMs over the redshift range 0.45-0.6.

Paper I presents an in-depth comparison between the mass estimates
from \citet{Maraston:2013} and those from the {\sc s82-mgc}. For our
three fiducial redshift bins, there is a mean offset between the two
stellar mass estimates, denoted
$\delta=\log_{10}(M_*^{\rm S82-MGC})-\log_{10}(M_*^{\rm M13})$. In the
range $11.3<\log_{10}(M_*^{\rm S82-MCG})<11.6$, $\delta=0.11$ dex for
$z_2$, $\delta=0.14$ dex for $z_3$, $\delta=0.10$ dex for $z_4$.

\begin{figure*}
\begin{center}
\includegraphics[width=16cm]{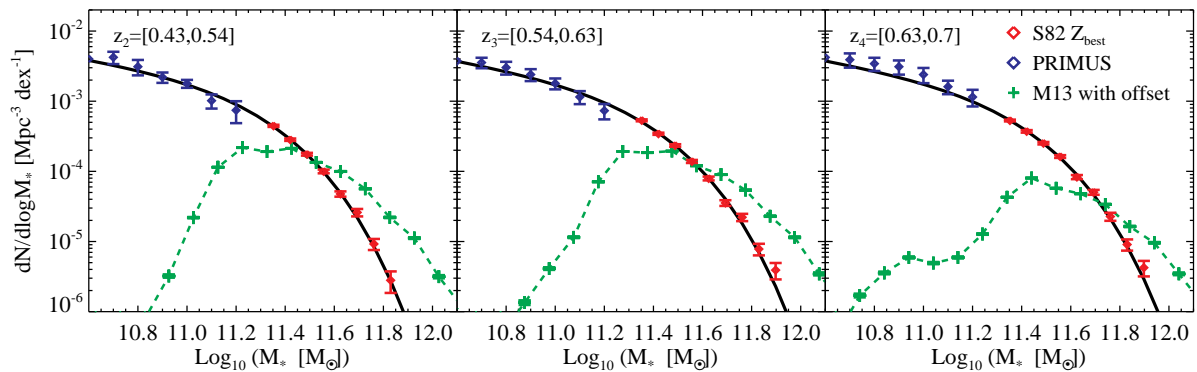}
\caption{Comparison between the total SMF derived in this work and the
  CMASS-only SMFs presented in \citet{Maraston:2013}. Green data
  points represent the \citet{Maraston:2013} SMFs after applying our
  estimate of the mean offset between the \citet{Maraston:2013} masses
  and {\sc s82-mgc} masses for each redshift bin (these offsets are of
  order 0.1-0.15 dex). The \citet{Maraston:2013} SMFs are computed for
  CMASS galaxies and do not include any completeness correction
  factors. }
\label{maraston}
\end{center}
\end{figure*}

With a better understanding of the CMASS sample now in hand, we
re-investigate the mass completeness of these CMASS SMFs from this
early BOSS analysis. Figure \ref{maraston} presents a comparison
between the total SMF derived in this paper with the CMASS SMFs
derived in \citet{Maraston:2013} (adjusted to our fiducial
cosmology). Green data points indicate the \citet{Maraston:2013} mass
functions after applying the mean offsets for each redshift bin. We
find a lower amplitude in our total SMF at the high-mass end compared
to \citet{Maraston:2013}. One possible explanation for this difference
is that the \citet{Maraston:2013} mass estimates may have a larger
scatter compared to the {\sc s82-mgc}, which are based on deeper
optical and NIR photometry. A larger mass error would cause the
\citet{Maraston:2013} SMFs to be inflated at the high-mass end
relative to those computed from the {\sc s82-mgc} (due to Eddington
bias).

By qualitatively comparing the CMASS mass functions with previous
analyses, \citet{Maraston:2013} provided a first estimate of the
completeness of the CMASS sample, tentatively concluding a rough mass
completeness of $\log_{10}(M_*/M_{\odot})=11.3$ at $z< 0.6$ and at
$\log_{10}(M_*/M_{\odot})=11.6$ at $z> 0.6$.  The analysis presented
here suggests that these original estimates were reasonable, although
somewhat optimistic.  A more accurate estimate is presented in Table
\ref{cmass_comp_table}.

\subsection{Halo Occupation Modeling from \citet{Shankar:2014} }
\citet{Shankar:2014} recently studied the high-mass slope and scatter
of the stellar-to-halo mass relation to $z=1$ using a variety of
different data sets. One component of their analysis is based on Halo
Occupation Distribution \citep[HOD,][and references
therein]{Berlind:2002} modeling of the projected two-point correlation
function of CMASS galaxies. This clustering analysis was performed in
two redshift bins: $0.4<z<0.6$ and $0.6<z<0.8$. A stellar mass cut of
$\log_{10}(M_*/M_{\odot})>11.5$ was applied to both samples. The
\citet{Maraston:2013} passive template mass catalog was adopted for
this analysis (Hong Guo, priv comm)\footnote{As opposed to
  creating a merged catalog using both the passive and the
  star-forming templates as in \citet{Maraston:2013}.}.

One underlying assumption of the HOD-type model employed by
\citet{Shankar:2014} is that stellar-mass selected threshold samples should
be mass complete. In this case, the central occupation function of
central galaxies is generally expected to be well described by a
traditional {\it erf} function that converges to unity at large
masses. If the samples under consideration are incomplete in mass,
then the amplitude and form of the central occupation function is
uncertain.

\begin{figure*}
\begin{center}
\includegraphics[width=17.5cm]{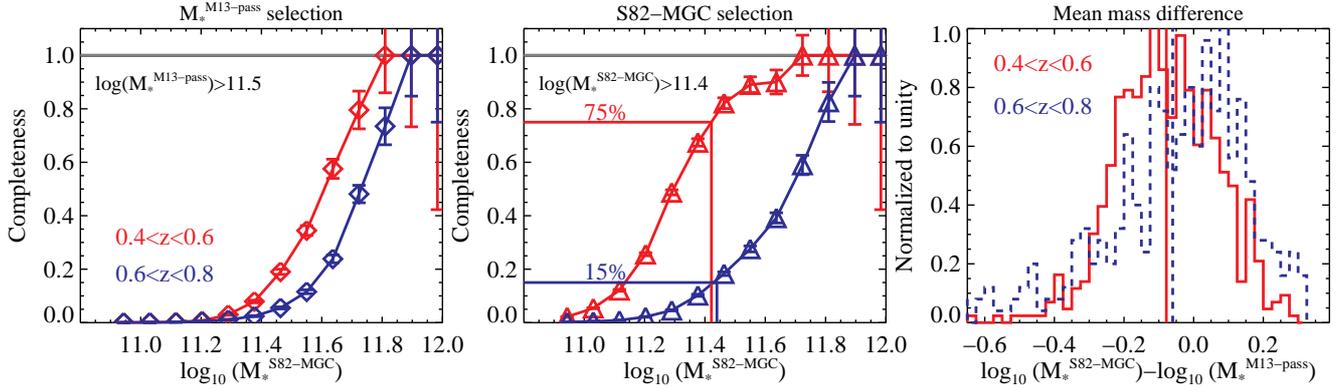}
\caption{Stellar mass completeness of the CMASS samples used in the
  \citet{Shankar:2014} analysis. Red diamonds correspond to the
  low-redshift sample and blue diamonds indicate the high-redshift
  sample. The x-axis represents the mass estimate from the {\sc
    s82-mgc}. Left panel: mass completeness evaluated using the actual
  \citet{Shankar:2014} sample but relative to {\sc s82-mgc}
  masses. This panel shows additional spread compared to a pure
  threshold sample due to scatter between the \citet{Maraston:2013}
  passive template masses and the {\sc s82-mgc} masses. Middle panel:
  mass completeness evaluated using a fixed {\sc s82-mgc} stellar mass
  cut. Right panel: distribution of mass differences between the
  \citet{Maraston:2013} passive template masses and the {\sc s82-mgc}
  masses. Vertical lines indicate the 50th percentile of the
  distribution. A mass cut of $\log_{10}(M_*^{\rm M13-pass})=11.5$
  corresponds to a mass cut of $\log_{10}(M_*)\sim11.43$ in the {\sc
    s82-mgc}. }
\label{shankar_comp}
\end{center}
\end{figure*}

With this possibility in mind, we evaluate the mass completeness of
the two samples employed by \citet{Shankar:2014} using the same
methodology as in Section \ref{sect_cmass_comp}. To estimate the
completeness, we recompute the total SMF within each of the two
\citet{Shankar:2014} redshift bins and show our results in Figure
\ref{shankar_comp}.

There are two aspects of this analysis worth emphasizing.  First, in
practice, all stellar mass estimates are imperfect and will have
scatter relative to the true underlying stellar mass. Hence, any
stellar mass ``threshold'' yields a sample with a more smoothly
varying selection function on the true $M_*$ distribution. We can
roughly estimate the magnitude of this effect by assuming that the
scatter between the \citet{Maraston:2013} passive template masses and
the {\sc s82-mgc} masses is similar to the scatter between either
estimate and the true underlying stellar mass. This effect will be
captured if we use the actual \citet{Shankar:2014} sample (i.e,
defined by a selection with respect to the \citealt{Maraston:2013}
masses) but evaluate the completeness using {\sc s82-mgc} masses. This
is shown in the left hand side of Figure \ref{shankar_comp}. The
samples no longer have a sharp boundary in mass but instead follow a
more smoothly varying completeness function.

Second, we can evaluate the completeness of a threshold cut on $M_*$
defined for one set of $M_*$ estimates by translating those thresholds
into $M_*$ used in this work.  The \citet{Shankar:2014} sample, for
example, was selected as $\log_{10}(M_*^{\rm M13-pass})>11.5$.
Comparing to the \citet{Maraston:2013} passive template masses used in
that work, we find a mean offset\footnote{Mass offsets will differ
  from those quoted in Section \ref{m13smf} because in one case a
  merged (star-forming plus passive) catalog was created, and in the
  other the passive template was adopted for all galaxies.} of
$\delta=-0.06$ dex and $\delta=-0.06$ for the low and high redshift
bins respectively compared to our masses from {\sc s82-mgc}.
Accounting for this mean offset, the fixed mass cut of
$\log_{10}(M_*^{\rm M13-pass})>11.5$ used by \citet{Shankar:2014}
corresponds to cuts of $\log_{10}(M_*)>11.43$ (low$-z$ bin) and
$\log_{10}(M_*)>11.44$ (high-$z$ bin) using our $M_*$ estimates.  At
these limits, CMASS is 75\% complete for the low-redshift
sample and only 15\% in the high-redshift sample (middle panel of
Figure \ref{shankar_comp}).

Finally, the right hand panel of Figure \ref{shankar_comp} displays
the difference between the \citet{Maraston:2013} passive template
masses and the {\sc s82-mgc} masses at these scales.  Both the offset
and relative scatter are apparent.

This exercise highlights the stellar mass incompleteness of these
samples and demonstrates that that caution needs to be taken when
studying CMASS. Understanding to what extend this may or may not have
an impact on the original \citet{Shankar:2014} analysis is beyond the
scope of this paper.

\subsection{Cosmological Analysis from \citet{Miyatake:2013} and \citet{More:2014}}
In two companion papers, \citet{Miyatake:2013} and \citet{More:2014}
present a joint analysis of the abundance, clustering, and
galaxy-galaxy lensing signal measured for CMASS sub-samples. Using a
HOD framework, they derive constraints on the high-mass end of the
stellar-to-halo mass relation and on the cosmological parameters
$\Omega_m$ and $\sigma_8$. \citet{Miyatake:2013} and \citet{More:2014}
consider three CMASS samples selected in the range $0.47<z<0.59$:
``Sample A'' with $\log_{10}(M_*/M_{\odot})>11.1$ (``cut 1''),
``Sample B'' with $\log_{10}(M_*/M_{\odot})>11.3$ (``cut 2''), and
``Sample C'' with $\log_{10}(M_*/M_{\odot})>11.4$ (``cut 3''). The
fiducial sample in the \citet{More:2014} cosmological analysis is the
A sample. Stellar masses are taken from the \citet{Maraston:2013}
passive-template catalog and assume a Kroupa IMF.

\begin{figure*}
\begin{center}
\includegraphics[width=17.5cm]{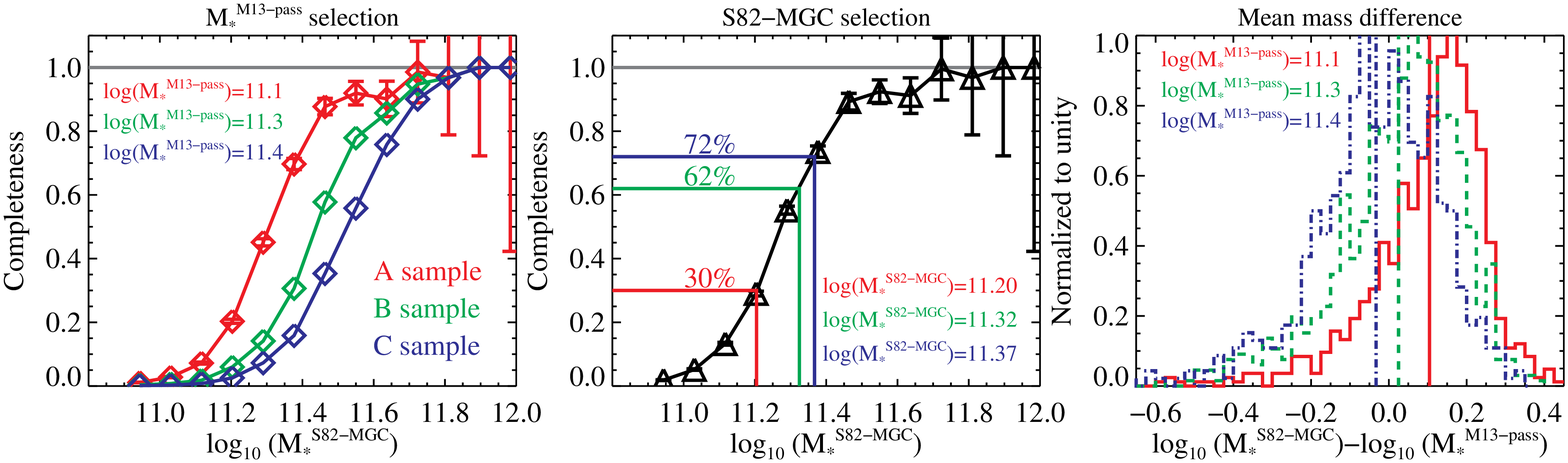}
\caption{Stellar mass completeness of the CMASS samples employed by
  \citet{Miyatake:2013} and \citet{More:2014}. The x-axis represents
  the mass estimate from the {\sc s82-mgc}. Left panel: mass
  completeness evaluated using the \citet{Miyatake:2013} and
  \citet{More:2014} samples but relative to {\sc s82-mgc} masses. This
  panel shows additional spread compared to a pure threshold sample
  due to scatter between the \citet{Maraston:2013} passive template
  masses and the {\sc s82-mgc} masses. Middle panel: mass completeness
  evaluated using a fixed {\sc s82-mgc} stellar mass cut. Right panel:
  distribution of mass differences between the \citet{Maraston:2013}
  passive template masses and the {\sc s82-mgc} masses. Vertical lines
  indicate the 50th percentile of the distribution. A mass cut of
  $\log_{10}(M_*^{\rm M13-pass})=11.1$ corresponds to a mass cut of
  $\log_{10}(M_*)\sim11.2$ in the {\sc s82-mgc}. }
\label{surhud_comp}
\end{center}
\end{figure*}

Using the same methodology as in the previous section, we compute the
mass completeness for these three samples.  The results are presented
in Figure \ref{surhud_comp}. Sample A is 30\% complete at cut 1,
sample B is 62\% complete at cut 2 and sample C is 72\% complete at
cut 3. The right hand panel shows the incompleteness function of these
samples evaluated using the original cuts. Again, the incompleteness
of the two samples relative to the true underlying mass probably lies
between the left and the middle panel of Figure \ref{surhud_comp}.

The \citet{More:2014} HOD analysis accounts for potential
incompleteness in the selection of CMASS galaxies at the low-mass end
when compared to a true stellar mass threshold sample. However, the
difficulty when working with incomplete samples is that galaxies that
are excluded from the sample are not a random population. For example,
the \citet{More:2014} incompleteness model assumes that the CMASS
selection corresponds to a random selection at fixed stellar and halo
mass, i.e., it assumes that the galaxies that are removed from the
sample live in similar halo environments as CMASS galaxies at fixed
stellar mass. However, as shown in Figure \ref{cmass_birth_parameter},
CMASS is color selected and preferentially selects certain regions of
color space at fixed stellar mass. In our companion paper, we
demonstrate that a model which accounts for the stellar mass
completeness of the CMASS sample, and reproduces the SMF and the two
point correlation functions, but assumes that color in uncorrelated
with environment at fixed stellar mass, fails to reproduce the
monopole and the quadrupole of the correlation function (Saito et
al. in preparation). Hence, the modeling of incomplete and
color-selected samples may not be straight-forward.

Again, a full evaluation of the impact of incompleteness on the
conclusions of \citet{Miyatake:2013} and \citet{More:2014} is beyond
the scope of this paper.  The better characterization of the mass and
color completeness of the BOSS samples presented here will make such
evaluations possible and improve future attempts to accurately model
the BOSS samples.

%%%%%%%%%%%%%%%%%%%%%%%%%%%%%%%%%%%%%%%%%%%%%%%%%%%%%%%%%%%%%%%%%%%%%%%%%%%%%%
%     CONCLUSIONS
%%%%%%%%%%%%%%%%%%%%%%%%%%%%%%%%%%%%%%%%%%%%%%%%%%%%%%%%%%%%%%%%%%%%%%%%%%%%%%

\section{Summary and Conclusions}\label{conclusions}

The BOSS survey has collected spectra for over one million galaxies at
$0.15<z<0.7$ over a volume of 15.3 Gpc$^3$ (9,376 deg$^2$) which
provides an opportunity to study the most massive galaxy populations
with vanishing sample variance. However, the BOSS sample selections
involve complex color cuts which are not necessarily optimized for
galaxy science. As a result, the selection function and stellar mass
completeness of these samples are poorly understood. Nonetheless,
given the large volumes and consequently large sample sizes at play,
these surveys have a tremendous potential to constrain the galaxy-halo
connection and to investigate the most massive galaxies in the universe
providing that the samples are well understood.

In this paper, we characterize the stellar mass completeness of the
BOSS samples with the goal of enabling future studies to better
utilize these samples for galaxy-formation science. We use data from
Stripe 82, which is roughly 2 magnitudes deeper than the single epoch
SDSS imaging, and construct a catalog of massive galaxies that is
complete to $\log_{10}(M_*/M_{\odot})>11.2$ at $z=0.7$. Using this
catalog, we empirically derive the stellar mass completeness of the
two main BOSS spectroscopic samples: the LOWZ sample at $0.15<z<0.43$
and the CMASS sample at $0.43<z<0.7$. We provide convenient fitting
formulas which can be used to estimate the completeness of each of
these samples as a function of stellar mass and redshift.

We demonstrate that CMASS is significantly impacted by mass
incompleteness and is 80\% complete at $\log_{10}(M_*/M_{\odot})>11.6$
only in the narrow redshift range $z=[0.51,0.61]$. At the mean
redshift of the CMASS sample, $\overline{z}=0.55$, CMASS is roughly
80\% complete at $\log_{10}(M_*/M_{\odot})=11.4$.  Based on these
considerations, referring to this sample in terms of ``constant mass''
should only be considered in loose terms. In contrast, we demonstrate
that LOWZ is 80\% complete at $\log_{10}(M_*/M_{\odot})>11.6$ over the
entire redshift range and 90\% complete at
$\log_{10}(M_*/M_{\odot})>11.5$ in the redshift range
$z=[0.18,0.29]$. At the mean redshift of the LOWZ sample,
$\overline{z}=0.29$, LOWZ is 80\% complete at
$\log_{10}(M_*/M_{\odot})=11.4$. Hence, our results counter the
conventional notion that CMASS is more complete at higher stellar
masses compared to LOWZ \citep[][]{Anderson:2014}.

Our results suggest an interesting redshift window for studying the
evolution of the most massive galaxies. The {\em combination} of LOW
and CMASS yields a spectroscopic sample that is 80\% complete at
$\log_{10}(M_*/M_{\odot})>11.6$ at $z<0.61$.

The values provided in this paper should be considered as {\em
  estimates}. As can be seen in Figures \ref{cmass_comp} and
\ref{lowz_comp}, the errors on the completeness at high stellar mass
are large due to the limited volume of Stripe 82. Upcoming wide area
surveys will be able to repeat this analysis with much higher
precision using several hundred to thousands of square degrees. The
values presented here are reported with respect to {\sc s82-mcg}
masses so offsets may need to be applied to translate these
completeness values to other mass estimates -- see Paper I for
details.

With upcoming surveys in mind, such as the HSC and Euclid surveys,
which will overlap with the BOSS footprint, we also characterize how
many supplementary galaxies with photometric redshifts will be needed
at any given stellar mass and redshift bin in order to construct mass
limited samples. A sample that is mass limited to
$\log_{10}(M_∗/M_{\odot}) > 11.6$ can be constructed at $z < 0.61$ by
supplementing the BOSS samples with photometric redshifts at the
$\sim$20\% level. At $\log_{10}(M_∗/M_{\odot})>11.0$, however,
spectroscopic samples need to be significantly supplemented by
photometric redshifts (at the 80\% level).

We use our methodology to evaluate the stellar mass completeness of
several specific samples that have been used in past work using BOSS
data. We demonstrate that previous work has sometimes over-estimated
the stellar mass completeness of BOSS samples and suggest that caution
needs to be taken when analyzing BOSS samples for these types of
studies.

The completeness estimates provided by this paper will enable future
studies to better utilize the BOSS samples for galaxy-formation and
cosmology science. A better understanding of the BOSS selection
functions will also enable the construction of improved mock catalogs
for the BOSS survey. Our companion paper presents improved mock
catalogs that account for the stellar mass completeness of the BOSS
CMASS sample as a function of redshift (Saito et al. in preparation).

\section*{Acknowledgements}

We thank Jean Coupon, Antonio Montero Dorta, Surhud More, Hironao
Miyatake, and Francesco Shankar for useful discussions while preparing
this paper. We are grateful to Hong Guo for kindly providing
catalogs. This work was supported by World Premier International
Research Center Initiative (WPI Initiative), MEXT, Japan. RT
acknowledges support from the Science and Technology Facilities
Council via an Ernest Rutherford Fellowship (grant number
ST/K004719/1). Funding for SDSS-III has been provided by the Alfred
P. Sloan Foundation, the Participating Institutions, the National
Science Foundation, and the U.S. Department of Energy Office of
Science. The SDSS-III web site is
\url{http://www.sdss3.org/}. SDSS-III is managed by the Astrophysical
Research Consortium for the Participating Institutions of the SDSS-III
Collaboration including the University of Arizona, the Brazilian
Participation Group, Brookhaven National Laboratory, Carnegie Mellon
University, University of Florida, the French Participation Group, the
German Participation Group, Harvard University, the Instituto de
Astrofisica de Canarias, the Michigan State/Notre Dame/JINA
Participation Group, Johns Hopkins University, Lawrence Berkeley
National Laboratory, Max Planck Institute for Astrophysics, Max Planck
Institute for Extraterrestrial Physics, New Mexico State University,
New York University, Ohio State University, Pennsylvania State
University, University of Portsmouth, Princeton University, the
Spanish Participation Group, University of Tokyo, University of Utah,
Vanderbilt University, University of Virginia, University of
Washington, and Yale University.

%-------------- APPENDIX -----------------------------------------------------
\appendix
\section{Catalogs}

Here we list the various links to the publicly available catalogs used
in this paper. 

$\bullet$ The original catalog used to target BOSS galaxies. The
target catalog for Stripe 82 is bosstarget-lrg-main007-collate.fits
and can be found at
\url{http://data.sdss3.org/sas/dr10/boss/target/main007/}. The target
selection flags are contained in the bitfield BOSS\_TARGET1. A
description of BOSS\_TARGET1 can be found at
\url{https://www.sdss3.org/dr10/algorithms/bitmask\_boss_target1.php}. Various
targets can be selected via (BOSS\_TARGET1 \&\&$ 2^i) > 0$ where $i$ is
a binary digit. In this paper, we are primarily concerned with the
samples GAL$\_$LOZ ($i=0$), GAL\_CMASS which corresponds to the CMASS
selection described in section \ref{bossdata}($i=1$), SDSS\_KNOWN
which corresponds to objects with Legacy objects ($i=6$), and
GAL\_CMASS\_ALL which includes GAL\_CMASS and the entire sparsely
sampled region ($i=7$). 

%;maskbits BOSS_TARGET1  0 GAL_LOZ          # low-z lrgs
%;maskbits BOSS_TARGET1  1 GAL_CMASS        # dperp > 0.55, color-mag cut
%;maskbits BOSS_TARGET1  3 GAL_CMASS_SPARSE # GAL_CMASS_COMM & (!GAL_CMASS) & (i < 19.9) sparsely sampled
%;maskbits BOSS_TARGET1  7 GAL_CMASS_ALL       # GAL_CMASS and the entire sparsely sampled region
%;maskbits BOSS_TARGET1  6  SDSS_GAL_KNOWN   # Matches to existing
%ZWARNING=0 SDSS galaxy spectra

$\bullet$ The Portsmouth stellar mass catalogs can be found at
\url{https://www.sdss3.org/dr10/spectro/galaxy\_portsmouth.php}.

$\bullet$ The DR10 large-scale structure catalogs
\citep[][]{Anderson:2014} that we use are
galaxy\_DR10v8\_CMASS\_South.fits and
galaxy\_DR10v8\_LOWZ\_South.fits. These catalogs can be found at
\url{http://data.sdss3.org/sas/dr10/boss/lss/} and the data model can
be found at \url{http://data.sdss3.org/datamodel/files/BOSS\_LSS\_REDUX/galaxy\_DR10v8\_SAMPLE\_NS.html}.

 %-------------- BIBLIO -------------------------------------------------------

\bibliographystyle{mnras}
\bibliography{mn-jour,all_refs}

\label{lastpage}

 %-----------------------------------------------------------------------------

\end{document}